\RequirePackage{ifpdf}
\ifpdf 
\documentclass[pdftex]{sigma}
\else
\documentclass{sigma}
\fi

\newcommand{\uim}{UV/IR mixing}
\newcommand{\nc}{non-com\-mu\-ta\-tive}
\newcommand{\etal}{et al.}
\newcommand{\figref}[1]{Fig.~\ref{#1}}			
\newcommand{\secref}[1]{Section~\ref{#1}}		
\newcommand{\co}[2]{\left[#1,#2\right]}					
\newcommand{\aco}[2]{\left\{#1,#2\right\}}				
\newcommand{\starco}[2]{\left[ #1\stackrel{\star}{,}#2\right] }		
\newcommand{\var}[2]{\frac{\d #1}{\d #2}}				
\newcommand{\pa}{\partial}						
\newcommand{\diff}[2]{\frac{\pa #1}{\pa #2}}				
\newcommand{\ri}{{\rm i}}						
\renewcommand{\k}{\tilde{k}}						
\newcommand{\p}{\tilde{p}}						
\newcommand{\q}{\tilde{q}}						
\newcommand{\Dt}{\widetilde{D}}						
\newcommand{\bc}{\bar{c}}						
\newcommand{\Act}{S}
\renewcommand{\a}{\alpha}
\renewcommand{\b}{\beta}

\renewcommand{\d}{\delta}
\newcommand{\e}{\epsilon}

\renewcommand{\th}{\theta}

\renewcommand{\l}{\lambda}
\newcommand{\m}{\mu}
\newcommand{\n}{\nu}

\renewcommand{\r}{\rho}
\newcommand{\s}{\sigma}
\renewcommand{\t}{\tau}

\renewcommand{\L}{\Lambda}

\newcommand{\W}{\Omega}

\newcommand{\inv}[1]{\frac{1}{#1}}				

\newcommand{\intk}{\int\limits_{-\infty}^{+\infty}\!\! \frac{{\rm d}^4k}{\left(2\pi\right)^4}}
\newcommand{\intx}{\int\! {\rm d}^4x}				
\newcommand{\ig}{{\rm i}g}
\renewcommand{\intk}{\int\!\!d^4k\,} 
\newcommand{\nov}[1]{\left(#1^2+\frac{\gamma^4}{#1^2}\right)} 	
\newcommand{\nok}{\left(k^2+\frac{a'^2}{\tilde{k}^2}\right)} 	
\newcommand{\bpsi}{\bar{\psi}}
\newcommand{\bB}{\bar{B}}
\newcommand{\oldB}{\mathcal{B}}
\newcommand{\ST}{\mathcal{B}_{S}}
\newcommand{\go}{\mathcal{G}}
\newcommand{\ag}{\bar{\mathcal{G}}}
\newcommand{\U}{\mathcal{U}^{(0)}}
\newcommand{\Ut}{\tilde{\mathcal{U}}^{(0)}}
\newcommand{\Uu}{\mathcal{U}^{(1)}}

\newcommand{\Q}{\mathcal{Q}}
\numberwithin{equation}{section}

\begin{document}

\allowdisplaybreaks

\renewcommand{\thefootnote}{$\star$}

\renewcommand{\PaperNumber}{037}

\FirstPageHeading

\ShortArticleName{Analysis of the Localized Non-Commutative $p^{-2}$ $U(1)$ Gauge Model}

\ArticleName{One-Loop Calculations and Detailed Analysis of the\\ Localized Non-Commutative $\boldsymbol{p^{-2}}$ $\boldsymbol{U(1)}$ Gauge Model\footnote{This paper is a
contribution to the Special Issue ``Noncommutative Spaces and Fields''. The
full collection is available at
\href{http://www.emis.de/journals/SIGMA/noncommutative.html}{http://www.emis.de/journals/SIGMA/noncommutative.html}}}

\Author{Daniel N. BLASCHKE~$^{\dag\ddag}$, Arnold ROFNER~$^\dag$ and Ren\'e I.P. SEDMIK~$^\dag$}

\AuthorNameForHeading{D.N. Blaschke, A. Rofner and R.I.P. Sedmik}

\Address{$^\dag$~Institute for Theoretical Physics,
Vienna University of Technology,\\
\hphantom{$^\dag$}~Wiedner Hauptstrasse 8-10, A-1040 Vienna, Austria}

\EmailD{\href{mailto:blaschke@hep.itp.tuwien.ac.at}{blaschke@hep.itp.tuwien.ac.at}, \href{mailto:arofner@hep.itp.tuwien.ac.at}{arofner@hep.itp.tuwien.ac.at},  \\ \hspace*{17.0mm}\href{mailto:sedmik@hep.itp.tuwien.ac.at}{sedmik@hep.itp.tuwien.ac.at}}

\Address{$^\ddag$~Faculty of Physics,
University of Vienna, Boltzmanngasse 5, A-1090 Vienna, Austria}

\ArticleDates{Received February 10, 2010, in f\/inal form April 23, 2010;  Published online May 04, 2010}

\Abstract{This paper carries forward a series of articles describing our enterprise to construct a gauge equivalent for the $\theta$-deformed {\nc} $\inv{p^2}$ model originally introdu\-ced by Gurau \etal~[{\it Comm.\ Math.\ Phys.} {\bf 287} (2009), 275--290]. It is shown that breaking terms of the form used by Vilar \etal~[{\it J.~Phys.~A: Math.\ Theor.} {\bf 43} (2010), 135401, 13~pages] and ourselves [{\it Eur.\ Phys.~J.~C: Part.\ Fields} {\bf62} (2009), 433--443] to localize the BRST covariant operator $\left(D^2\theta^2D^2\right)^{-1}$ lead to dif\/f\/iculties concerning renormalization. The reason is that this dimensionless operator is invariant with respect to any symmetry of the model, and can be inserted to arbitrary power. In the present article we discuss explicit one-loop calculations, and analyze the mechanism the mentioned problems originate from.}


\Classification{81T13; 81T15; 81T75}

\renewcommand{\thefootnote}{\arabic{footnote}}
\setcounter{footnote}{0}

\section{Introduction}

\label{sec:introduction}

Tackling the infamous UV/IR mixing problem~\cite{Minwalla:1999, Susskind:2000} plaguing Moyal-deformed QFTs has been one of the main research interests in the f\/ield for almost a decade (see \cite{Tanasa:2008d,Rivasseau:2007a,Douglas:2001} for reviews of the topic). It is accepted on a broad basis that non-commutativity necessitates additional terms in the action to reobtain renormalizability. Several interesting approaches have been worked out~\cite{Grosse:2003,Grosse:2008a}, and proofs of renormalizability have been achieved mainly by utilizing Multiscale Analysis (MSA)~\cite{Rivasseau:2005a,Rivasseau:2005b}, or formally in the matrix base~\cite{Grosse:2004b}.

In the line of these developments Gurau \etal~\cite{Gurau:2009} introduced a term of type $\phi\star\frac{a}{\square}\phi$ into the Lagrangian which, in a natural way, provides a counter term for the inevitable $\inv{p^2}$ divergence inherently tied to the deformation of the product. In this way the theory is altered in the infrared region which breaks the {\uim} and renders the theory renormalizable. This latter fact has been proven up to all orders by the authors using MSA. Motivated by the inherent translation invariance and simplicity of this model (referred to as $\inv{p^2}$ model), a thorough study of the divergence structure and explicit renormalization at one-loop level~\cite{Blaschke:2008b}, as well as a~computation of the beta functions~\cite{Tanasa:2008a} have been carried out.

In the present article we work on Euclidean $\mathbb{R}_\th^4$ with the Moyal-deformed product (also referred to as `star product') $\starco{x_\m}{x_\n} \equiv  x_{\mu} \star x_{\nu} -x_{\nu} \star x_{\mu} = \ri \th_{\mu \nu}$ of regular commuting coordinates $x_\mu$. In the simplest case, the real parameters $\th_{\m\n}=-\th_{\n\m}$ form the block-diagonal tensor
\begin{gather*}
( \th_{\mu\nu} )
=\th\left(\begin{array}{cccc}
0&1&0&0\\
-1&0&0&0\\
0&0&0&1\\
0&0&-1&0
\end{array}\right) ,  \qquad \text{with} \ \ \th \in \mathbb{R}  ,
\end{gather*}
obeying the practical relation $\theta_{\m\r}\theta_{\r\n}=-\th^2\d_{\m\n}$, where $\dim \th=-2$.
With these def\/initions we use the abbreviations $\tilde{v}_\m \equiv \th_{\m\n}v_\n$ for vectors $v$ and $\tilde{M} \equiv \th_{\m\n}M_{\m\n}$ for matrices $M$.

Further research focused on the generalization of the scalar $\inv{p^2}$ model to $U_{\star}(1)$ gauge theory\footnote{Notice, that the star product modif\/ies the initial $U(1)$ algebra in a way that
it becomes non-Abelian. Hence, we call the resulting algebra $U_\star(1)$.} which was f\/irst proposed in~\cite{Blaschke:2008a} yielding the action
\begin{gather}
 \Act =\Act_{\text{inv}}[A]+\Act_{\text{gf}}[A,b,c,\bc]\nonumber\\
\phantom{\Act}{} = \intx\Big[\inv{4}F_{\m\n}\star F_{\m\n}\!+F_{\m\n}\star\inv{D^2\Dt^2}\star F_{\m\n}\Big]\!+\!\intx \Big[b\star\partial\cdot A-\frac{\a}{2}b\star b - \bc\star\partial_\m D_\m c\Big],\!\!\!
 \label{eq:act_old_complete}
\end{gather}
with the usual gauge boson $A_\m$, ghost and antighost f\/ields $c$ and $\bc$ respectively, the Lagrange multiplier f\/ield $b$ implementing the gauge f\/ixing, and a real $U_{\star}(1)$ gauge parameter $\a$. The antisymmetric f\/ield strength tensor $F_{\m\n}$ and the covariant derivative $D_\m$ are def\/ined by
\begin{gather*}
 F_{\m\n}=\partial_\m A_\n -\partial_\n A_\m -\ri g \starco{A_{\m}}{A_{\n}}, \qquad \text{and} \qquad D_{\m}\varphi= \partial_\m\varphi-\ri g \starco{A_\m}{\varphi},
\end{gather*}
for arbitrary $\varphi$. The non-local term
\begin{align}
\Act_{\text{nloc}}=\intx\, F_{\m\n}\star\inv{D^2\Dt^2}\star F_{\m\n},
\label{eq:act_nl_old}
\end{align}
implements the damping mechanism of the $\inv{p^2}$ model by Gurau~\etal~\cite{Gurau:2009} in a gauge covariant way. It has been described in~\cite{Blaschke:2008b} that the new operator can only be interpreted in a physically sensible way if it is cast into an inf\/inite series which, however, corresponds to an inf\/inite number of gauge vertices. A f\/irst attempt to localize the new operator by introducing a real valued auxiliary tensor f\/ield~\cite{Blaschke:2009a} led to additional degrees of freedom. However, this was considered to be dissatisfactory. Following the ideas of Vilar \etal~\cite{Vilar:2009} we enhanced our approach by coupling gauge and auxiliary sectors via complex conjugated pairs of f\/ields together with associated pairs of ghosts in such a way, that BRST doublet structures were formed \cite{Blaschke:2009b}. Such a mechanism has already been applied successfully for the Gribov--Zwanziger action of QCD~\cite{Zwanziger:1989,Zwanziger:1993,Dudal:2008} where a~similar damping mechanism is applied.

Starting from a recapitulation of our recently presented localized model in \secref{sec:loc_p2inv} we give explicit one-loop calculations in \secref{sec:one_loop}, and undertake the attempt of one-loop renor\-ma\-lization. Subsequently, the results and their implications for higher loop orders are analyzed in \secref{sec:higher_loop_calc}, and f\/inally we give a concluding discussion of the lessons learned in \secref{sec:discussion}.

\section[The localized $p^{-2}$ $U(1)$ gauge model]{The localized $\boldsymbol{\inv{p^2}}$ $\boldsymbol{U(1)}$ gauge model}

\label{sec:loc_p2inv}

\subsection{Review: the construction of the model}\label{sec:review}

As mentioned in \secref{sec:introduction}, the non-local term of the action \eqref{eq:act_old_complete} leads to an inf\/inite number of vertices: it formally consists of the inverse of covariant derivatives acting on f\/ield strength tensors, and therefore stands for an inf\/inite power series (cf.~\cite{Blaschke:2008a}) making explicit calculations impossible. Considering only the f\/irst few orders of this power series is not an option as this would destroy gauge invariance. Yet, the present problem can be circumvented by the localization of the term under consideration.
In this sense, in a f\/irst approach described in~\cite{Blaschke:2009a}, the introduction of an additional real antisymmetric f\/ield $\oldB_{\m\n}$ of mass dimension two led to the following localized version of the non-local term \eqref{eq:act_nl_old}:
\begin{gather}
\Act_{\text{nloc}}\to\Act_{\text{loc}}=\intx\big[a'\oldB_{\mu\nu}\star F_{\mu\nu}-\oldB_{\mu\nu}\star \Dt^2D^2\star \oldB_{\mu\nu}\big].
\label{eq:gauge_act_damping_replacement}
\end{gather}
However, the $\oldB_{\m\n}$-f\/ield appears to have its own dynamical properties leading to new physical degrees of freedom which can only be avoided if the new terms in the action are written as an exact BRST variation. In order for such a mechanism to work, further unphysical f\/ields are required.

Following the ideas of Vilar \etal~\cite{Vilar:2009}, the localized action \eqref{eq:gauge_act_damping_replacement} was further developed in~\cite{Blaschke:2009b} by replacing $\oldB_{\m\n}$ with a complex conjugated pair of f\/ields ($B_{\m\n}$, $\bB_{\m\n}$) and by the introduction of an additional pair of ghost and antighost f\/ields $\psi_{\m\n}$ and $\bpsi_{\m\n}$ (all of mass dimension~1), thus leading to
\begin{gather}\label{act-loc}
\Act_{\text{loc}} =\intx\left[\frac{\l}{2}\left(B_{\m\n}+
\bB_{\m\n}\right)F_{\m\n}-\mu^2\bB_{\m\n}D^2\Dt^2B_{\m\n}+\mu^2\bpsi_{\m\n}D^2\Dt^2\psi_{\m\n}\right].
\end{gather}
In this expression, as well as throughout the remainder of this section, all f\/ield products are~con\-sidered to be star products. The new parameters $\l$ and $\mu$ both have mass dimension 1 and replace the former dimensionless parameter $a'$ by $a'=\l/\m$. The proof of the equivalence between the non-local action \eqref{eq:act_nl_old} and equation~\eqref{act-loc} can be found in~\cite{Blaschke:2009b}. With the addition of a f\/ixing term to the action one has BRST invariance, and for simplicity, we choose the Landau gauge
\begin{gather*}
\Act_{\phi\pi}=\intx\left(b\partial^\m A_\m-\bc\partial^\m D_\m c\right).
\end{gather*}
The BRST transformation laws for the f\/ields read:
\begin{alignat}{3}
 &sA_\mu=D_\mu c,\qquad && sc=\ri g{c}{c} ,& \nonumber\\
 &s\bc=b,     \qquad                                                  && sb=0 ,&  \nonumber\\
& s\bpsi_{\mu\nu}=\bB_{\m\n}+\ri g\aco{c}{\bpsi_{\m\n}}, \qquad                && s\bB_{\m\n}=\ri g\co{c}{\bB_{\m\n}}, & \nonumber\\
& sB_{\m\n}=\psi_{\m\n}+\ri g\co{c}{B_{\m\n}}, \qquad                   && s\psi_{\m\n}=\ri g\aco{c}{\psi_{\m\n}},& \nonumber\\
&s^2\varphi=0\quad \forall\; \varphi\in\left\{A_\mu,b,c,\bc,B_{\m\n},\bB_{\m\n},\psi_{\m\n},\bpsi_{\m\n}\right\}.\quad&&&\label{eq:BRST_all_fields}
\end{alignat}
With \eqref{eq:BRST_all_fields} one can see that the localized part of the action can be written as the sum of a BRST exact and a so-called soft breaking term:
\begin{gather*}
\Act_{\text{loc}}=\intx\left[s\left(\frac{\l}{2}\bpsi_{\m\n}F^{\m\n}
-\mu^2\bpsi_{\m\n}D^2\Dt^2B^{\m\n}\right)+\frac{\l}{2}B_{\m\n}F^{\m\n}\right],
\end{gather*}
where
\begin{gather}\label{eq:act_break_wo_source}
\Act_{\text{break}}=\intx\, \frac{\l}{2}B_{\m\n}F^{\m\n}, \qquad \text{with }\quad s\Act_{\text{break}} =\intx\, \frac{\l}{2}\psi_{\m\n}F^{\m\n}.
\end{gather}
As discussed in detail in \cite{Blaschke:2009b}, the breaking is considered to be soft, since the mass dimension of the f\/ield dependent part is $<D=4$ and the term only modif\/ies the infrared regime of the model. As has been shown by Zwanziger~\cite{Zwanziger:1993} terms of this type therefore do not spoil renormalizability. In order to restore BRST invariance in the UV region (as is a prerequisite for the application of algebraic renormalization) an additional set of sources{\samepage
\begin{alignat*}{3}
&s \bar{Q}_{\m\n\a\b}=\bar{J}_{\m\n\a\b}+\ri g \aco{c}{\bar{Q}_{\m\n\a\b}},\qquad && s \bar{J}_{\m\n\a\b}=\ri g \co{c}{\bar{J}_{\m\n\a\b}}, &
\nonumber\\
&s Q_{\m\n\a\b}=J_{\m\n\a\b}+\ri g \aco{c}{Q_{\m\n\a\b}},\qquad  && s J_{\m\n\a\b}=\ri g \co{c}{J_{\m\n\a\b}}, & 
\end{alignat*}}

\noindent
is introduced, and coupled to the breaking term which then takes the (BRST exact) form
\begin{gather*}
\Act_{\text{break}} =\intx s\left(\bar{Q}_{\m\n\a\b}B^{\m\n}F^{\a\b}\right)
 =\intx \left(\bar{J}_{\m\n\a\b}B^{\m\n}F^{\a\b}-\bar{Q}_{\m\n\a\b}\psi^{\m\n}F^{\a\b}\right) .
\end{gather*}
Equation~\eqref{eq:act_break_wo_source} is reobtained if the sources  $\bar{Q}$ and $\bar{J}$ take their `physical values'
\begin{alignat}{3}
& \bar{Q}_{\m\n\a\b}\big|_{\text{phys}}=0,\qquad  && \bar{J}_{\m\n\a\b}\big|_{\text{phys}}=\frac{\l}{4}\left(\d_{\m\a}\d_{\n\b}-\d_{\m\b}\d_{\n\a}\right),& \nonumber\\
& Q_{\m\n\a\b}\big|_{\text{phys}}=0,\qquad && J_{\m\n\a\b}\big|_{\text{phys}}=\frac{\l}{4}\left(\d_{\m\a}\d_{\n\b}-\d_{\m\b}\d_{\n\a}\right).& \label{JQ-phys}
\end{alignat}
Note that the Hermitian conjugate of the counter term $\Act_{\text{break}}$ in equation~\eqref{act-loc} (i.e.\ the term $\intx\bB_{\m\n}F^{\m\n}$) may also be coupled to external sources which, however, is not required for BRST invariance but restores Hermiticity of the action:
\begin{gather*}
\frac{\l}{2}\intx \, \bB_{\m\n}F^{\m\n}\ \longrightarrow \ \intx\,s\big(J_{\m\n\a\b}\bpsi^{\m\n}F^{\a\b}\big)=\intx\,J_{\m\n\a\b}\bB^{\m\n}F^{\a\b}.
\end{gather*}

Including external sources $\W^\phi$, $\phi\in\{A,c,B,\bB,\psi,\bpsi,J,\bar{J},Q,\bar{Q}\}$ for the non-linear BRST transformations the complete action with Landau gauge $\pa^\m A_\m=0$ and general $Q/\bar{Q}$ and $J/\bar{J}$ reads:
\begin{gather}
\Act =\Act_{\text{inv}}+\Act_{\phi\pi}+\Act_{\text{new}}+\Act_{\text{break}}+\Act_{\text{ext}} ,\qquad \text{with}\nonumber\\
\Act_{\text{inv}} =\intx\inv{4}F_{\m\n}F_{\m\n} ,\nonumber\\
\Act_{\phi\pi} =\intx\,s\left(\bc\,\pa_\m A_\m\right)=\intx\left(b\,\pa_\m A_\m-\bc\,\pa_\m D_\m c\right) ,
\nonumber\\
\Act_{\text{new}} =\intx\,s\big(J_{\m\n\a\b}\bpsi_{\m\n}F_{\a\b}-\mu^2\bpsi_{\m\n}D^2\Dt^2B_{\m\n}\big)\nonumber\\
\phantom{\Act_{\text{new}}}{}
=\intx\big(J_{\m\n\a\b}\bB_{\m\n}F_{\a\b}-\mu^2\bB_{\m\n}D^2\Dt^2B_{\m\n}+\mu^2\bpsi_{\m\n}D^2\Dt^2\psi_{\m\n}\big),
\nonumber\\
\Act_{\text{break}} =\intx\,s\big(\bar{Q}_{\m\n\a\b}B_{\m\n}F_{\a\b}\big)
=\intx\big(\bar{J}_{\m\n\a\b}B_{\m\n}F_{\a\b}-\bar{Q}_{\m\n\a\b}\psi_{\m\n}F_{\a\b}\big),\nonumber\\
\Act_{\text{ext}} =\intx\left(\W^A_\m D_\m c+\ig\, \W^c c c+\W^{B}_{\m\n}\big(\psi_{\m\n}+\ig\co{c}{B_{\m\n}}\big) +\ig\,\W^{\bB}_{\m\n}\co{c}{\bB_{\m\n}}\right.\nonumber\\
 \phantom{\Act_{\text{ext}} =}{}
 +\ig\, \W^\psi_{\m\n}\aco{c}{\psi_{\m\n}}+ \W^{\bpsi}_{\m\n}\big(\bB_{\m\n}+\ig \aco{c}{\bpsi_{\m\n}}\big)+\W^{Q}_{\m\n\a\b}\big( J_{\m\n\a\b}+\ig \aco{c}{Q_{\m\n\a\b}}\big)  \nonumber\\
\left.
\phantom{\Act_{\text{ext}} =}{}
+\ig\, \W^J_{\m\n\a\b}\co{c}{J_{\m\n\a\b}}+\W^{\bar{Q}}_{\m\n\a\b}\big( \bar{J}_{\m\n\a\b}+\ig \aco{c}{\bar{Q}_{\m\n\a\b}}\big) +\ig\, \W^{\bar{J}}_{\m\n\a\b}\co{c}{\bar{J}_{\m\n\a\b}}\right).\!\!
\label{eq:act_complete}
\end{gather}
Table~\ref{tab:field_prop} summarizes properties of the f\/ields and sources contained in the model~\eqref{eq:act_complete}.

\begin{table}[!th]\centering
\caption{Properties of f\/ields and sources.}
\label{tab:field_prop}
\vspace{1mm}

\begin{tabular}{l c c c c c c c c c c c}
\hline
\hline
\rule[12pt]{0pt}{0.1pt}
Field       & $A_\m$ & $c$ & $\bc$ & $B_{\m\n}$ & $\bB_{\m\n}$ & $\psi_{\m\n}$ & $\bpsi_{\m\n}$ & $J_{\a\b\m\n}$ & $\bar{J}_{\a\b\m\n}$ & $Q_{\a\b\m\n}$ & $\bar{Q}_{\a\b\m\n}$\\[2pt]
\hline
$g_\sharp$  &    0   &  1  &   $-1$  &     0      &    0         &        1      &      $-1$        &   0            &       0              &  $-1$            &  $-1$\tsep{1pt}\\
Mass dim.   &    1   &  0  &   2   &     1      &    1         &        1      &       1        &   1            &       1              &  1             &  1\\
Statistics  &    b   &  f  &   f   &     b      &    b         &        f      &       f        &   b            &       b              &  f             &  f\\
\hline
\rule[14pt]{0pt}{0.1pt}
Source      & $\W^A_\m$ & $\W^c$ & $b$ & $\W^B_{\m\n}$ & $\W^{\bB}_{\m\n}$ & $\W^{\psi}_{\m\n}$ & $\W^{\bpsi}_{\m\n}$ & $\W^J_{\a\b\m\n}$ & $\W^{\bar{J}}_{\a\b\m\n}$ &$\W^Q_{\a\b\m\n}$ & $\W^{\bar{Q}}_{\a\b\m\n}$\\[2pt]
\hline
$g_\sharp$  &   $-1$   &  $-2$ &   0   &    $-1$      &   $-1$         &       $-2$      &       0        &  $-1$            &      $-1$              & 0             &  0\tsep{1pt}\\
Mass dim.   &    3   &  4  &   2   &     3      &    3         &        3      &       3        &   3            &       3              &  3            &  3\\
Statistics  &    f   &  b  &   b   &    f     &    f         &        b      &      b        &   f            &       f              &  b            &  b\\
\hline
\hline
\end{tabular}
\end{table}

Notice that the mass $\mu$ is a physical parameter despite the fact that the variation of the action $\diff{\Act}{\mu^2}=s\big(\bpsi_{\m\n}D^2\Dt^2B^{\m\n}\big)$ yields an exact BRST form. Following the argumentation in~\cite{Baulieu:2009} this is a consequence of the introduction of a soft breaking term. For vanishing Gribov-like parameter $\l$ the contributions to the path integral of the $\mu$ dependent sectors of $\Act_{\text{new}}$ in \eqref{eq:act_complete} cancel each other. If $\l\neq0$ one has to consider the additional breaking term which couples the gauge f\/ield $A_\mu$ to the auxiliary f\/ield $B_{\mu\nu}$ and the associated ghost $\psi_{\m\n}$. This mixing is ref\/lected by the appearance of $a'=\l/\mu$ in the damping factor $\nok$ featured by all f\/ield propagators \eqref{eq:prop_aa}--\eqref{eq:prop_bb} below.

\subsection{Feynman rules}

\subsubsection{Propagators}
From the action \eqref{eq:act_complete} with $J/\bar J$ and $Q/\bar Q$ set to their physical values given by \eqref{JQ-phys} one f\/inds the propagators
\begin{subequations}
\begin{gather}
G^{\bc c}(k) =-\inv{k^2} ,\label{eq:prop_cc}\\
G^{\bpsi\psi}_{\m\n,\r\s}(k) =\frac{\left(\d_{\m\r}\d_{\n\s}-\d_{\m\s}\d_{\n\r}\right)}{2\mu^2k^2\k^2}\,,\label{eq:prop_psipsi}\\
G^{AA}_{\m\n}(k) =\inv{\nok}\left(\d_{\m\n}-\frac{k_\m k_\n}{k^2}\right) ,\label{eq:prop_aa}\\
G^{AB}_{\m,\r\s}(k) =\frac{\ri a'}{2\mu}\frac{\left(k_\r\d_{\m\s}-k_\s\d_{\m\r}\right)}{k^2\k^2\nok}=G^{A\bB}_{\m,\r\s}(k)=-G^{\bB A}_{\r\s,\m}(k) ,\label{eq:prop_ab}\\
G^{\bB B}_{\m\n,\r\s}(k) =\frac{-1}{2\mu^2k^2\k^2}\left[\d_{\m\r}\d_{\n\s}-\d_{\m\s}\d_{\n\r}-a'^2\frac{k_\m k_\r\d_{\n\s}{+}k_\n k_\s\d_{\m\r}{-}k_\m k_\s\d_{\n\r}{-}k_\n k_\r\d_{\m\s}}{2k^2\k^2\nok}\right]\!,\!\!\!\!\label{eq:prop_bbarb}\\
G^{BB}_{\m\n,\r\s}(k) =\frac{a'^2}{2\m^2k^2\k^2}\left[\frac{k_\m k_\r\d_{\n\s}+k_\n k_\s\d_{\m\r}-k_\m k_\s\d_{\n\r}-k_\n k_\r\d_{\m\s}}{2k^2\k^2\nok}\right]=G^{\bB\bB}_{\m\n,\r\s}(k) ,\label{eq:prop_bb}
\end{gather}
\end{subequations}
where the abbreviation $a'\equiv\l/\mu$ is used. Notice, that they obey the following symmetries and relations:
\begin{subequations}
\label{proprel}
\begin{gather}
 G^{AB}_{\m,\r\s}(k) =G^{A\bB}_{\m,\r\s}(k)=-G^{BA}_{\r\s,\m}(k)=-G^{\bB A}_{\r\s,\m}(k),\label{prop-rel_a}\\
 G^{\phi}_{\m\n,\r\s}(k) =-G^{\phi}_{\n\m,\r\s}=-G^{\phi}_{\m\n,\s\r}(k)=G^{\phi}_{\n\m,\s\r}(k),\qquad
 \text{for} \ \ \phi \mathrel{\in}\{\bpsi\psi,\bB B, BB, \bB \bB\},\!\!\! \\
 2k^2\k^2 G^{AB}_{\r,\m\n}(k) =\ri\frac{a'}{\m}\left(k_\m G^{AA}_{\r\n}(k)-k_\n G^{AA}_{\r\m}(k)\right),\label{prop-rel_c}\\
  \inv{\m^2}\left(\d_{\mu\rho}\d_{\nu\s}-\d_{\mu\s}\d_{\nu\rho}\right) =\ri\frac{a'}{\m}\left(k_\m G^{BA}_{\r\s,\n}(k)-k_\n G^{BA}_{\r\s,\m}(k)\right)-2k^2\k^2 G^{B\bB}_{\mu\nu,\rho\s}(k),\\
  0 =\ri\frac{a'}{\m}\left(k_\m G^{BA}_{\r\s,\n}(k)-k_\n G^{BA}_{\r\s,\m}(k)\right)-2k^2\k^2G^{BB}_{\mu\nu,\rho\s}(k),\label{prop-rel_f}\\
  G^{B\bB}_{\m\n,\r\s}(k) =G^{\bpsi\psi}_{\m\n,\r\s}(k)+G^{BB}_{\m\n,\r\s}(k).
\end{gather}
\end{subequations}

\subsubsection{Vertices}
The action \eqref{eq:act_complete} leads to 13 tree level vertices whose rather lengthy expressions are listed in Appendix~\ref{app:vertices}. One immediately f\/inds the following vertex relation:
\begin{gather*}
\widetilde{V}^{\bpsi \psi (n\times A)}_{\mu\nu,\rho\s,\xi_1\ldots\xi_n}(q_1, q_2, k_{\xi_1},\ldots, k_{\xi_n}) =
-\widetilde{V}^{\bB B (n\times A)}_{\mu\nu,\rho\s,\xi_1\ldots\xi_n}(q_1, q_2, k_{\xi_1},\ldots, k_{\xi_n}),
\end{gather*}
i.e.\ all vertices with one $B$, one $\bB$ and an arbitrary number of $A$ legs have exactly the same form as the ones with one $\psi$, one $\bpsi$ and an arbitrary number of $A$ legs. This is due to the fact that the $\bpsi \psi nA$ and $\bB B nA$ vertices stem from terms in the action which are of the same structure, and are thus equal in their form.

Finally, the vertices obey the following additional relations:
\begin{gather*}
 \widetilde{V}^{\bpsi \psi (n\times A)}_{\mu\nu,\rho\s,\xi_1\ldots\xi_n}(q_1, q_2, k_{\xi_1},\ldots, k_{\xi_n}) =-\widetilde{V}^{\psi \bpsi (n\times A)}_{\rho\s,\mu\nu,\xi_1\ldots\xi_n}(q_2,q_1, k_{\xi_1}, \ldots, k_{\xi_n})\nonumber\\
\qquad {} =-\widetilde{V}^{\bpsi \psi (n\times A)}_{\n\m,\r\s,\xi_1\ldots\xi_n}(q_1,q_2, k_{\xi_1},\ldots, k_{\xi_n})
 =-\widetilde{V}^{\bpsi \psi (n\times A)}_{\m\n,\s\r,\xi_1\ldots\xi_n}(q_1,q_2, k_{\xi_1},\ldots, k_{\xi_n}),
\end{gather*}
and
\begin{gather*}
 \widetilde{V}^{\bB B (n\times A)}_{\mu\nu,\rho\s,\xi_1\ldots\xi_n}(q_1, q_2, k_{\xi_1},\ldots, k_{\xi_n}) =+\widetilde{V}^{B \bB (n\times A)}_{\rho\s,\mu\nu,\xi_1\ldots \xi_n}(q_2,q_1, k_{\xi_1},\ldots, k_{\xi_n})\\  
 =-\widetilde{V}^{\bB B (n\times A)}_{\n\m,\r\s,\xi_1\ldots\xi_n}(q_1,q_2, k_{\xi_1},\ldots, k_{\xi_n})
 =-\widetilde{V}^{\bB B (n\times A)}_{\m\n,\s\r,\xi_1\ldots\xi_n}(q_1,q_2, k_{\xi_1},\ldots, k_{\xi_n}),\quad
\text{for}  \  n \in {1,2,3,4} .\nonumber
\end{gather*}

\subsection{Symmetries}

Before moving on to explicit one-loop calculations, let us brief\/ly discuss the symmetries of our action equation~\eqref{eq:act_complete}. The Slavnov--Taylor identity is given by
\begin{gather}
 \oldB(\Act)=\intx\Bigg[\var{\Act}{\W^A_\m}\var{\Act}{A_\m}+\var{\Act}{\W^c}\var{\Act}{c}+ b \var{\Act}{\bc}+\var{\Act}{\W^B_{\m\n}}\var{\Act}{B_{\m\n}}+\var{\Act}{\W^{\bB}_{\m\n}}\var{\Act}{\bB_{\m\n}} \nonumber\\
\phantom{\oldB(\Act)=}{}+\var{\Act}{\W^\psi_{\m\n}}\var{\Act}{\psi_{\m\n}}
+\var{\Act}{\W^{\bpsi}_{\m\n}}\var{\Act}{\bpsi_{\m\n}}+\var{\Act}{\W^{Q}_{\m\n\a\b}}\var{\Act}{Q_{\m\n\a\b}}
+\var{\Act}{\W^J_{\m\n\a\b}}\var{\Act}{J_{\m\n\a\b}}\nonumber\\
 \phantom{\oldB(\Act)=}{}+\var{\Act}{\W^{\bar{Q}}_{\m\n\a\b}}\var{\Act}{\bar{Q}_{\m\n\a\b}}
+\var{\Act}{\W^{\bar{J}}_{\m\n\a\b}}\var{\Act}{\bar{J}_{\m\n\a\b}}\Bigg]=0.\label{mod2-SL-id}
\end{gather}
Furthermore we have the gauge f\/ixing condition
\begin{gather*}
\var{\Act}{b}=\pa_\m A_\m=0,
\end{gather*}
the ghost equation
\begin{gather*}
\go(\Act)=\pa_\m\var{\Act}{\Omega^A_\m}+\var{\Act}{\bc}=0,
\end{gather*}
and the antighost equation
\begin{gather*}
\ag(\Act)=\intx \var{\Act}{c}=0.
\end{gather*}

Following the notation of~\cite{Vilar:2009} the identity associated to the BRST doublet structure is given~by
\begin{gather*}
\Uu_{\a\b\m\n}(\Act)=\intx\Bigg(\bB_{\a\b}\var{\Act}{\bpsi_{\m\n}}+\W^{\bpsi}_{\m\n}\var{\Act}{\W^{\bB}_{\a\b}}
+\psi_{\m\n}\var{\Act}{B_{\a\b}}-\W^B_{\a\b}\var{\Act}{\W^{\psi}_{\m\n}} \nonumber\\
\phantom{\Uu_{\a\b\m\n}(\Act)=}{}
+J_{\m\n\r\s}\var{\Act}{Q_{\a\b\r\s}}+\W^{Q}_{\a\b\r\s}\var{\Act}{\W^J_{\m\n\r\s}}
+\bar{J}_{\a\b\r\s}\var{\Act}{\bar{Q}_{\m\n\r\s}}+\W^{\bar{Q}}_{\m\n\r\s}\var{\Act}{\W^{\bar{J}}_{\a\b\r\s}}\Bigg)=0.
\end{gather*}
Note that the f\/irst two terms of the second line,
\[
\intx\left( J_{\m\n\r\s}\var{\Act}{Q_{\a\b\r\s}}+\W^{Q}_{\a\b\r\s}\var{\Act}{\W^J_{\m\n\r\s}}\right)=0,
\]
constitute a symmetry by themselves. These terms stem from the insertion of conjugated f\/ield partners $J$ and $Q$ for $\bar{J}$ and $\bar{Q}$, respectively, which are not necessarily required as discussed above in~\secref{sec:review}.

Furthermore, we have the linearly broken symmetries $\U$ and $\Ut$:
\begin{gather*}
\U_{\a\b\m\n}(\Act)=-\Theta^{(0)}_{\a\b\m\n}=-\Ut_{\a\b\m\n}(\Act),
\end{gather*}
with
\begin{gather*}
\U_{\a\b\m\n}(\Act)=\intx\Bigg[B_{\a\b}\var{\Act}{B_{\m\n}}-\bB_{\m\n}\var{\Act}{\bB_{\a\b}}
-\W^B_{\m\n}\var{\Act}{\W^B_{\a\b}}+\W^{\bB}_{\a\b}\var{\Act}{\W^{\bB}_{\m\n}} \nonumber\\
\phantom{\U_{\a\b\m\n}(\Act)=}{} +J_{\a\b\r\s}\var{\Act}{J_{\m\n\r\s}}-\bar{J}_{\m\n\r\s}\var{\Act}{\bar{J}_{\a\b\r\s}}
-\W^J_{\m\n\r\s}\var{\Act}{\W^J_{\a\b\r\s}}+\W^{\bar{J}}_{\a\b\r\s}\var{\Act}{\W^{\bar{J}}_{\m\n\r\s}}\Bigg],\nonumber\\
\Ut_{\a\b\m\n}(\Act)= \intx\Bigg[\psi_{\a\b}\var{\Act}{\psi_{\m\n}}-\bpsi_{\m\n}\var{\Act}{\bpsi_{\a\b}}
-\W^{\psi}_{\m\n}\var{\Act}{\W^{\psi}_{\a\b}}+\W^{\bpsi}_{\a\b}\var{\Act}{\W^{\bpsi}_{\m\n}} \nonumber\\
\phantom{\Ut_{\a\b\m\n}(\Act)=}{} +Q_{\a\b\r\s}\var{\Act}{Q_{\m\n\r\s}}-\bar{Q}_{\m\n\r\s}\var{\Act}{\bar{Q}_{\a\b\r\s}}
-\W^Q_{\m\n\r\s}\var{\Act}{\W^Q_{\a\b\r\s}}+\W^{\bar{Q}}_{\a\b\r\s}\var{\Act}{\W^{\bar{Q}}_{\m\n\r\s}}\Bigg],\nonumber\\
\Theta^{(0)}_{\a\b\m\n}=\intx\left[\bB_{\m\n}\W^{\bpsi}_{\a\b}-\psi_{\a\b}\W^B_{\m\n}
+\bar{J}_{\m\n\r\s}\W^{\bar{Q}}_{\a\b\r\s}-J_{\a\b\r\s}\W^{Q}_{\m\n\r\s}\right].
\end{gather*}
The above relations would, if applicable, form the starting point for the algebraic renormalization procedure. In order to assure the completeness of the set of symmetries it has to be assured that the algebra generated by them closes. From the Slavnov--Taylor identity \eqref{mod2-SL-id} one derives the linearized Slavnov operator
\begin{gather*}
 \ST= \intx\Bigg[\var{\Act}{\W^A_\m}\var{\ }{A_\m}+\var{\Act}{A_\m}\var{\ }{\W^A_\m}+\var{\Act}{c}\var{\ }{\W^c}+\var{\Act}{\W^c}\var{\ }{c}+ b \var{\Act}{\bc}+\var{\Act}{\W^B_{\m\n}}\var{\ }{B_{\m\n}}+\var{\Act}{B_{\m\n}}\var{\ }{\W^B_{\m\n}}\nonumber\\
\phantom{\ST=}{} +\var{\Act}{\W^{\bB}_{\m\n}}\var{\ }{\bB_{\m\n}}+\var{\Act}{\bB_{\m\n}}\var{\ }{\W^{\bB}_{\m\n}}+\var{\Act}{\W^\psi_{\m\n}}\var{\ }{\psi_{\m\n}}+\var{\Act}{\psi_{\m\n}}\var{\ }{\W^\psi_{\m\n}}+\var{\Act}{\W^{\bpsi}_{\m\n}}\var{\ }{\bpsi_{\m\n}}+\var{\Act}{\bpsi_{\m\n}}\var{\ }{\W^{\bpsi}_{\m\n}}\nonumber\\
\phantom{\ST=}{}+\var{\Act}{\W^{Q}_{\m\n\a\b}}\var{\ }{Q_{\m\n\a\b}}+\var{\Act}{Q_{\m\n\a\b}}\var{\ }{\W^{Q}_{\m\n\a\b}}+\var{\Act}{\W^J_{\m\n\a\b}}\var{\ }{J_{\m\n\a\b}}+\var{\Act}{J_{\m\n\a\b}}\var{\ }{\W^J_{\m\n\a\b}}\nonumber\\
\phantom{\ST=}{} +\var{\Act}{\W^{\bar{Q}}_{\m\n\a\b}}\var{\ }{\bar{Q}_{\m\n\a\b}}+\var{\Act}{\bar{Q}_{\m\n\a\b}}\var{\ }{\W^{\bar{Q}}_{\m\n\a\b}}+\var{\Act}{\W^{\bar{J}}_{\m\n\a\b}}\var{\ }{\bar{J}_{\m\n\a\b}}+\var{\Act}{\bar{J}_{\m\n\a\b}}\var{\ }{\W^{\bar{J}}_{\m\n\a\b}}\Bigg].
\end{gather*}
Furthermore, the $\U$ and $\Ut$ symmetries are combined to def\/ine the operator $\Q$ as
\begin{gather*}
\Q\equiv\d_{\a\m}\d_{\b\n}\big(\U_{\a\b\m\n}+\Ut_{\a\b\m\n}\big).
\end{gather*}
Notice that the action is invariant under $\Q$, i.e.\ $\Q(\Act)=0$ because of $\U_{\a\b\m\n}(\Act)=-\Ut_{\a\b\m\n}(\Act)$.

Having def\/ined the operators $\ST$, $\ag$, $\Q$ and $\Uu$ we may derive the following set of graded commutators:
\begin{alignat*}{4}
& \aco{\ag}{\ag}=0,\qquad & & \aco{\ST}{\ST}=0,\qquad  & &
\{\ag,\ST\}=0, & \nonumber\\
& \co{\ag}{\Q}=0, \qquad & & \co{\Q}{\Q}=0, \qquad & &
\big\{\ag,\Uu_{\m\n\a\b}\big\}=0,  & \nonumber\\
&
\big\{\ST,\Uu_{\m\n\a\b}\big\}=0,\qquad & &
\big\{\Uu_{\m\n\a\b},\Uu_{\m'\n'\a'\b'}\big\}=0,
\qquad  & &
\big[\Uu_{\m\n\a\b},\Q\big]=0,
& \nonumber\\
& \co{\ST}{\Q}=0,&& && &
\end{alignat*}
which shows that the algebra of symmetries closes.

Having derived the symmetry content of the model, we would now be ready to apply the method of Algebraic Renormalization (AR). The latter requires locality which, however, is not given in the present case and generally for all {\nc} QFTs, due to the inherent non-locality of the star product. Hence, before the application of AR it would be required to establish the foundations of this method also for {\nc} theories. For a detailed discussion we would like to refer to our recent article~\cite{Blaschke:2009c}.

\section{One-loop calculations}
\label{sec:one_loop}

In this section we shall present the calculations relevant for the one-loop correction to the gauge boson propagator. Due to the existence of the mixed propagators $G^{\text{AB}}$, $G^{\text{A}\bB}$, and their mirrored counterparts, the two point function $\langle A_\m A_\n\rangle$ receives contributions not only from graphs with external gauge boson legs, but also from those featuring external $B$ and/or $\bB$ f\/ields.

In the following (i.e.~in Sections~\ref{sec:1loop_vacpol}--\ref{sec:1loop_BbB}), we will present a detailed analysis of all truncated two-point functions relevant for the calculation of the full one-loop $AA$-propagator. Every type of correction, being characterized by its amputated external legs (i.e.~$A$, $B$ or $\bB$), is discussed in a separate subsection. Finally, in \secref{sec:1loop_analysis} the dressed $AA$-propagator and the attempt for its one-loop renormalization will be given explicitly.

\subsection{Vacuum polarization}
\label{sec:1loop_vacpol}

\begin{figure}[t]
 \centering
 \includegraphics[scale=0.8]{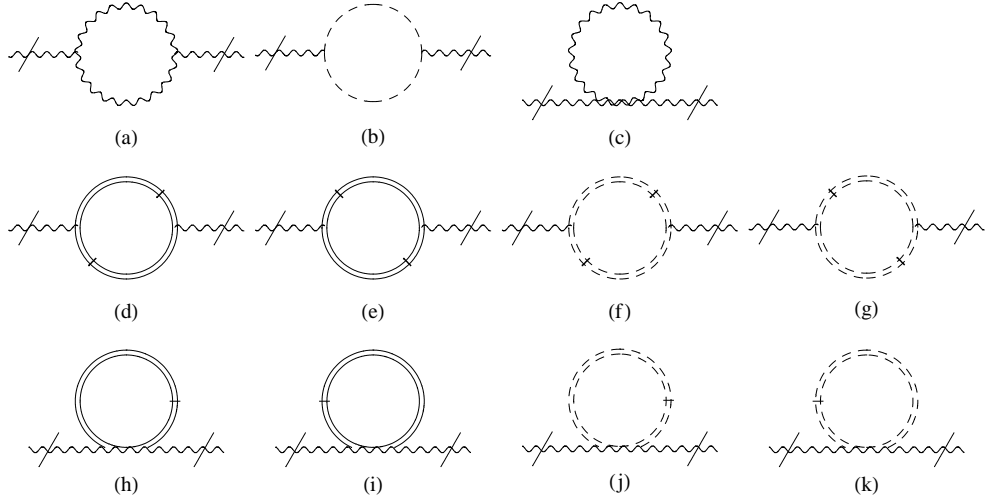}
 \caption{One loop corrections for the gauge boson propagator.}
 \label{fig:1loop_vacpol_all}
\end{figure}

The model \eqref{eq:act_complete} gives rise to 23 graphs contributing to the two-point function $G^{AA}_{\m\n}(p)$. Omitting convergent expressions, there are 11 graphs left depicted in \figref{fig:1loop_vacpol_all}.
Being interested in the divergent contributions one can apply the expansion~\cite{Blaschke:2009a}
\begin{gather}
\Pi_{\mu\nu}= \intk\, \mathcal{I}_{\mu\nu}(p,k) \sin^2\left(\tfrac{k\p}{2}\right)\approx
 \intk\sin^2\left(\tfrac{k\p}{2}\right)\bigg\{\mathcal{I}_{\mu\nu}(0,k)+p_\rho\left[\partial_{p_\rho}\mathcal{I}_{\mu\nu}(p,k)\right]_{p=0}\nonumber\\
\phantom{\Pi_{\mu\nu}=}{} +\frac{p_\rho p_\s}{2}\left[\partial_{p_\rho}\partial_{p_\s}\mathcal{I}_{\mu\nu}(p,k)\right]_{p=0}
+\mathcal{O}\left(p^3\right)\bigg\},\label{eq:1-l_expansion}
\end{gather}
where the integrand $\mathcal{I}_{\mu\nu}(p,k)$ has been separated from the phase factor in order to keep the regularizing ef\/fects in the non-planar parts due to rapid oscillations for large $k$. Summing up the contributions of the graphs in \figref{fig:1loop_vacpol_all} and denoting the result at order~$i$ for the planar (p) part by $\Pi_{\m\n}^{(i),\text{p}}$, one is left with
\begin{gather*}
 \Pi_{\m\n}^{(0),\text{p}}(p)= \frac{g^2}{16 \pi ^2} \L^2 \d _{\mu\n}\left(-10 s_{\rm c}-96 s_{\rm h}-96 s_{\rm j}+12 s_{\rm a}+s_{\rm b}+96 s_{\rm d}+96 s_{\rm f}\right) =0, \\ 
\Pi_{\m\n}^{(2),\text{p}}(p)= -\frac{1}{3} \frac{g^2}{16 \pi ^2} \Big[\d _{\mu\nu} p^2\left(22 s_{\rm a}+s_{\rm b}+48(s_{\rm d}+s_{\rm f})\right)\nonumber\\
 \phantom{\Pi_{\m\n}^{(2),\text{p}}(p)=}{}
 +2 p_{\mu} p_{\nu} \left(72 (s_{\rm h}+s_{\rm j})-8 s_{\rm a}+s_{\rm b}-96(s_{\rm d}+s_{\rm f})\right)\Big] {\rm K}_0\left(2 \sqrt{\frac{M^2}{\L ^2}}\right)\nonumber\\
 \phantom{\Pi_{\m\n}^{(2),\text{p}}(p)}{}= -\frac{5 g^2}{12 \pi ^2} \left(p^2 \d _{\mu\nu}-p_{\mu} p_{\nu}\right) K_0\left(2 \sqrt{\frac{M^2}{\L ^2}}\right) \nonumber\\
\phantom{\Pi_{\m\n}^{(2),\text{p}}(p)}{} \approx -\frac{5 g^2}{24 \pi ^2} \left(p^2 \d _{\mu\nu}-p_{\mu} p_{\nu}\right) \ln\left(\frac{\L^2}{M^2}\right)+\text{f\/inite},
\end{gather*}
where the symmetry factors in Table~\ref{tab:1loop_vacpol_sf} have been inserted and the approximation
\[
\mathrm{K}_0(x)\underset{x\ll1}{\approx}\ln\tfrac{2}{x}-\gamma_E+\mathcal{O}\left(x^2\right),
\]
for the modif\/ied Bessel function ${\rm K}_0$ can be utilized for small arguments, i.e.\ vanishing regulator cutof\/fs\footnote{The cutof\/fs are introduced via a factor $\exp\left[-M^2\alpha-\inv{\Lambda^2\alpha}\right]$ to regularize parameter integrals $\int_0^\infty{\rm d}\alpha$. See~\cite{Blaschke:2009a} for a more extensive description of the mathematical details underlying these computations.} $\Lambda\to \infty$ and $M\to0$. Finally, $\gamma_E$ denotes the Euler--Mascheroni constant. Note that the f\/irst order vanishes identically due to an odd power of $k$ in the integrand which leads to a~cancellation under the symmetric integration over the momenta.

\begin{table}[t]
 \centering
 \caption{Symmetry factors for the one loop vacuum polarization (where the factor $(-1)$ for fermionic loops has been included).}
 \begin{tabular}{l r | l r | l r}
\hline
 \hline
 $s_{\rm a}$ &$\inv{2}$ & $s_{\rm e}$ &  1& $s_{\rm i}$ & 1\tsep{2pt}\\
 $s_{\rm b}$ & $-1$ 	& $s_{\rm f}$ & $-1$& $s_{\rm j}$ & $-1$\\
 $s_{\rm c}$ &$\inv{2}$ & $s_{\rm g}$ & $-1$& $s_{\rm k}$ & $-1$\\
 $s_{\rm d}$ &1 	& $s_{\rm h}$ &  1&       &      \\
\hline
\hline
 \end{tabular}
 \label{tab:1loop_vacpol_sf}
\end{table}

Of particular interest is the non-planar part (np) which for small $p$ results to:
\begin{subequations}
\begin{gather}
 \Pi_{\m\n}^{(0),\text{np}}(p)= \frac{g^2}{4\pi ^2\p^2} \Big[\d _{\mu\nu}\left(96(s_{\rm h}+s_{\rm j}- s_{\rm d}- s_{\rm f})-12 s_{\rm a}-s_{\rm b}+10 s_{\rm c}\right)\nonumber\\*
 \phantom{\Pi_{\m\n}^{(0),\text{np}}(p)=}{}
 -2 \frac{\p_{\mu} \p_{\nu}}{\p^2} \left(48(s_{\rm h}+ s_{\rm j})-96 (s_{\rm d}+s_{\rm f})-12 s_{\rm a}-s_{\rm b}+2 s_{\rm c}\right)\Big]
= \frac{2g^2}{\pi^2}\frac{\p_\m \p_\n}{(\p^2)^2} , \label{eq:1loop_AA_o0_np}\\
   \Pi_{\m\n}^{(2),\text{np}}(p)= \frac{g^2}{48\pi^2\p^2} \bigg\{2 \th ^2 p_\mu p_\nu p^2\left(72 (s_{\rm h}+s_{\rm j})-8 s_{\rm a}+s_{\rm b}- 96(s_{\rm d}+ s_{\rm f})\right) \mathop{{\rm K}_0}\left(\sqrt{M^2 \p^2}\right)\nonumber\\ \phantom{\Pi_{\m\n}^{(2),\text{np}}(p)=}{}
   +\sqrt{\frac{\p^2}{M^2}}p^2\bigg[\sqrt{\frac{\p^2}{M^2}}\left(22 s_{\rm a}+s_{\rm b}+48(s_{\rm d}+s_{\rm f})\right)M^2 \d_{\mu\nu} \mathop{{\rm K}_0}\left(\sqrt{M^2 \p^2}\right)\nonumber\\
\phantom{\Pi_{\m\n}^{(2),\text{np}}(p)=}{}
+2 M^2 \left(13 s_{\rm a}+s_{\rm b}+120(s_{\rm d}+s_{\rm f})\right) \p_{\mu} \p_\n \mathop{{\rm K}_1}\left(\sqrt{M^2 \p^2}\right)\nonumber\\
\phantom{\Pi_{\m\n}^{(2),\text{np}}(p)=}{}
-3 \sqrt{\frac{M^2}{\p^2}} \left(16 s_{\rm a}+s_{\rm b}+96(s_{\rm d}+ s_{\rm f})\right) \p_\m \p_\n\bigg]\bigg\}\nonumber\\
\phantom{\Pi_{\m\n}^{(2),\text{np}}(p)}{}= -\frac{g^2}{48\pi^2} \bigg[\p_\m\p_\n\bigg(\frac{21}{\th^2}-11 p^2 \sqrt{\frac{M^2}{\p^2}}\mathop{{\rm K}_1}\left(\sqrt{M^2\p^2}\right)\bigg)\nonumber\\
\phantom{\Pi_{\m\n}^{(2),\text{np}}(p)=}{}
-10 \mathop{{\rm K}_0}\left(\sqrt{M^2 \p^2}\right) \left(p^2\d_{\m\n} -p_\m p_\n \right)\bigg].
 \label{eq:1loop_AA_o2_np}
\end{gather}
\end{subequations}
Considering the limit $\p^2\to0$ rectif\/ies application of the approximation
\[
\mathrm{K}_1(x)\underset{x\ll1}{\approx}\tfrac{1}{x}
+\tfrac{x}{2}\left(\gamma_E-\tfrac{1}{2}+\ln\tfrac{x}{2}\right)+\mathcal{O}\left(x^2\right),
\]
which reveals that the second order is IR f\/inite (which is immediately clear from the fact that the terms of lowest order in $p$ are $\mathcal{O}\left(p^2\right)$), apart from a $\ln (M^2)$-term which cancels in the sum of planar and non-planar contributions. Hence, collecting all divergent terms one is left with (in the limit $M\to0$ and $\L\to\infty$),
\begin{gather}\label{eq:complete-1loop-sum}
\Pi_{\mu\nu}(p) =\frac{2g^2}{\pi^2}\frac{\p_\m \p_\n}{(\p^2)^2}-\lim\limits_{\L\to\infty}\frac{5 g^2}{24 \pi ^2} \left(p^2 \delta _{\mu\nu}-p_{\mu} p_{\nu}\right) \ln\left(\Lambda^2\right)+\text{f\/inite terms},
\end{gather}
which is independent of the IR-cutof\/f $M$. As expected\footnote{In fact, equation~\eqref{eq:complete-1loop-sum} qualitatively resembles the result of the ``na\"ive'' {\nc} gauge model discussed e.g.\ in~\cite{Hayakawa:1999, Armoni:2000xr,Ruiz:2000,Blaschke:2005b}. The dif\/ferent numerical factor in front of the logarithmic UV divergence is a~consequence of the contribution of additional f\/ields in the current model.}, equation~\eqref{eq:complete-1loop-sum} exhibits a quadratic IR divergence in $\p^2$ and a logarithmic divergence in the cutof\/f $\L$. Furthermore, the transversality condition $p_\mu\Pi_{\mu\nu}(p)=0$ is fulf\/illed, which serves as a consistency check for the symmetry factors.

\subsection[Corrections to the $AB$ propagator]{Corrections to the $\boldsymbol{AB}$ propagator}
\label{sec:1loop_AB}

\begin{figure}[t]
 \centering
 \includegraphics[scale=0.8]{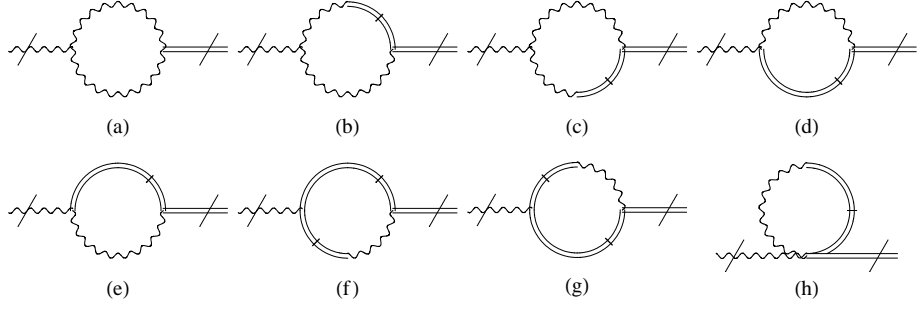}
 \caption{One loop corrections for $\langle A_\m B_{\n_1\n_2} \rangle$ (with amputated external legs).}
 \label{fig:1loop_AB}
\end{figure}
\begin{table}[t]
 \centering
 \caption{Symmetry factors for the graphs depicted in \figref{fig:1loop_AB}.}
 \vspace{1mm}

 \begin{tabular}{l r | l r }
\hline
 \hline
 $(a)$ & 1/2 &  $(e)$ & 1\tsep{2pt}\\
 $(b)$ & 1 & $(f)$ & 1 \\
 $(c)$ & 1 &  $(g)$ & 1 \\
 $(d)$ & 1 &  $(h)$ & 1\\
\hline
\hline
 \end{tabular}
 \label{tab:1loop_AB_div}
\end{table}

The action \eqref{eq:act_complete} gives rise to eight divergent graphs with one external $A_\m$ and one $B_{\m\n}$ which are depicted in~\figref{fig:1loop_AB}.
Applying an expansion of type~\eqref{eq:1-l_expansion} for small external momenta $p$ and summing up the divergent contributions of all graphs (all orders of an expansion similar to equation~\eqref{eq:1-l_expansion}) one ends up with,
\begin{gather*}
\Sigma_{\m1,\n1 \n2}^{\text{p,AB}}(p)=-\frac{3\ri g^2}{32 \pi^2} \lambda  \left(p_{\nu 1} \d_{\m1 \n2}-p_{\n2} \d_{\m1 \n1}\right) \mathop{{\rm K}_0}\left(2 \sqrt{\frac{M^2}{\Lambda^2}}\right)+{\rm f\/inite}, \nonumber\\
\Sigma_{\m1,\n1 \n2}^{\text{np,AB}}(p)= \frac{3\ri g^2}{32 \pi^2} \lambda \mathop{{\rm K}_0}\left(\sqrt{M^2 \p^2}\right) \left(p_{\nu1} \delta_{\mu1 \nu2}-p_{\nu2} \delta_{\mu1 \nu1}\right)+{\rm f\/inite}.
\end{gather*}
Approximating the Bessel functions as in \secref{sec:1loop_vacpol} and summing up planar and non-planar parts one f\/inds the expression
\begin{gather*}
\Sigma_{\m1,\n1\n2}^{\text{AB}}(p)= \frac{3\ri g^2}{32 \pi^2} \lambda \left(p_{\nu 1} \d_{\m1 \n2}-p_{\n2} \d_{\m1 \n1}\right)\left(\ln\Lambda+\ln|\p|\right)+{\rm f\/inite},
\end{gather*}
where the IR cutof\/f $M$ has cancelled, and which shows a logarithmic divergence for $\Lambda\to \infty$.

Due to the symmetry between $B$ and $\bB$ in the sense that both have identical interactions with the gauge f\/ield, it is obvious that $\Sigma_{\m1,\n1\n2}^{\rm AB}\equiv\Sigma_{\m1,\n1\n2}^{{\rm A}\bB}$ and as implied by equation~\eqref{prop-rel_a} it also holds that $\Sigma_{\m1\m2,\n1}^{\rm BA}\equiv-\Sigma_{\n1,\m1\m2}^{\rm AB}$.

\subsection[Corrections to the $BB$ propagator]{Corrections to the $\boldsymbol{BB}$ propagator}
\label{sec:1loop_BB}

\begin{figure}[t]
 \centering
 \includegraphics[scale=0.8]{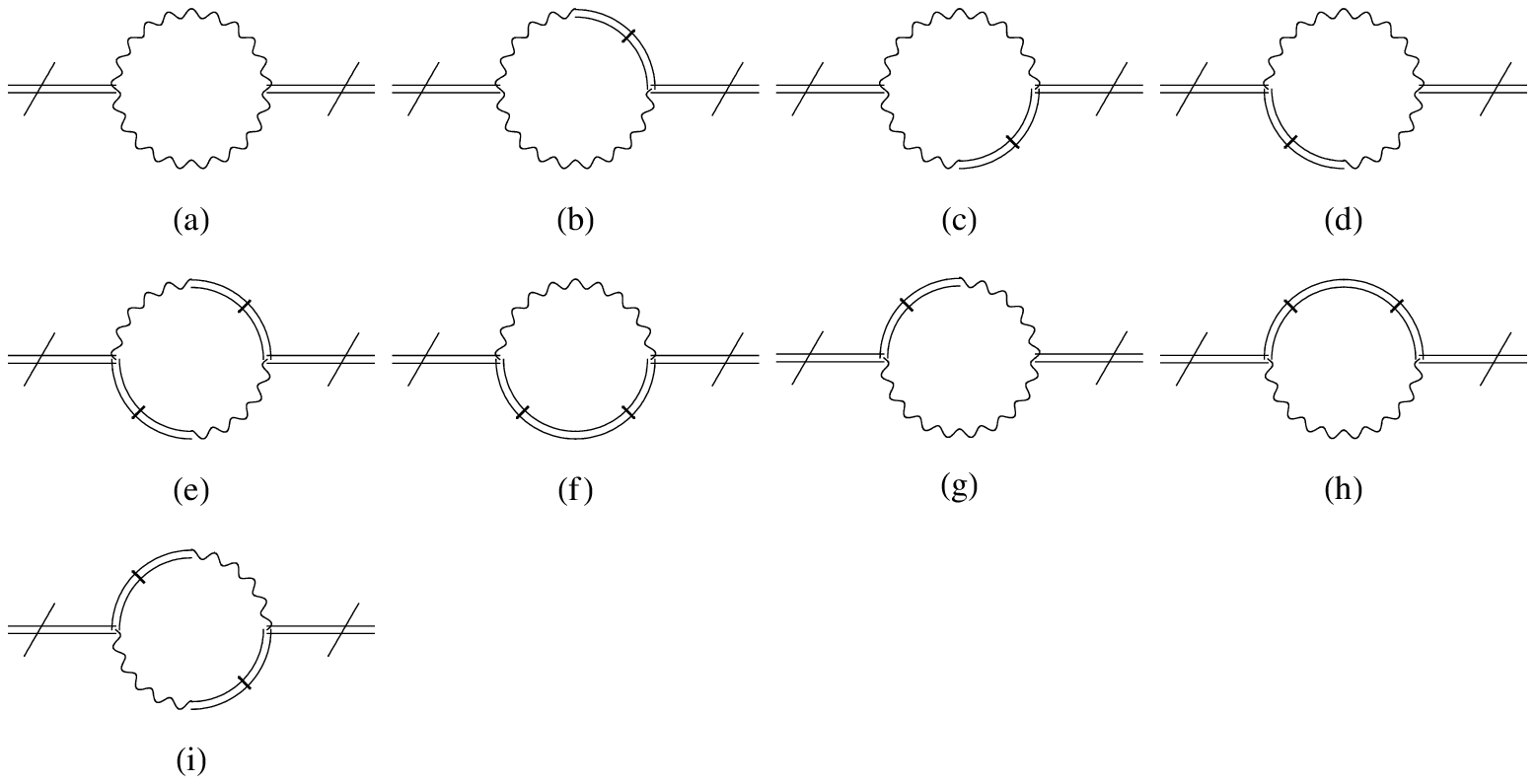}
 \caption{One loop corrections for $\langle  B_{\m_1\m_2} B_{\n_1\n_2} \rangle$ (with amputated external legs).}
 \label{fig:1loop_BB}
\end{figure}
\begin{table}[t]
 \centering
 \caption{Symmetry factors for the graphs depicted in \figref{fig:1loop_BB}.}
 \vspace{1mm}

 \begin{tabular}{l c | l c | l c}
\hline
 \hline
$(a)$ & 1/2 &  $(d)$ & 1 &  $(g)$ & 1\tsep{2pt}\\
$(b)$ & 1 &  $(e)$ & 1 &  $(h)$ & 1\\
$(c)$ & 1 &  $(f)$ & 1 &  $(i)$ & 1\\
\hline
\hline
 \end{tabular}
 \label{tab:1loop_BB_div}
\end{table}

The set of divergent graphs contributing to $\langle B_{\m1\m2}B_{\n1\n2} \rangle$ consists of those depicted in \figref{fig:1loop_BB}.
Making an expansion of type \eqref{eq:1-l_expansion} for small external momenta $p$ and summing up the contributions of all nine graphs yields
\begin{gather*}
\Sigma_{\m1\m2,\n1 \n2}^{\text{p,BB}}(p)= \frac{g^2 \lambda ^2}{32 \pi^2} \left(\d _{\m1 \n1} \d_{\m2 \n2}-\d _{\m2 \n1} \d _{\m1 \n2}\right)  \mathop{{\rm K}_0}\left(2 \sqrt{\frac{M^2}{\L ^2}}\right)+{\rm f\/inite}, \nonumber\\
\Sigma_{\m1\m2,\n1 \n2}^{\text{np,BB}}(p)= \frac{g^2 \lambda ^2}{64 \pi^2} \bigg(\frac{\d_{\m1\n2} \p_{\m2}\p_{\n1}-\d_{\m1 \n1}\p_{\m2}\p_{\n2}-\d_{\mu 2 \n 2} \p_{\m1} \p_{\n1}+\d_{\m2\n1}\p_{\m1}\p_{\n2}}{\p^2}\nonumber\\
 \phantom{\Sigma_{\m1\m2,\n1 \n2}^{\text{np,BB}}(p)= }{}
+2  \mathop{{\rm K}_0}\left(\sqrt{M^2 \p^2}\right) \left(\d_{\m1 \n2} \d_{\m2 \n1}-\d_{\m1 \n1} \d _{\m2 \n2}\right)\bigg)+{\rm f\/inite},
\end{gather*}
for the planar/non-planar part, respectively.
Approximating the Bessel functions as in \secref{sec:1loop_vacpol} reveals cancellations of contributions depending on $M$ in the f\/inal sum. Hence, the divergent part boils down to
\begin{gather}
\Sigma_{\m1\m2,\n1\n2}^{\text{BB}}(p)= \frac{g^2 \l^2}{64 \pi^2}\left(\d_{\m1 \n1} \d_{\m2 \n2}-\d_{\m2 \n1} \d_{\m1 \n2}\right)\left(\ln\L^2+\ln\p^2\right)+{\rm f\/inite},
\label{eq:1loop_BB_final}
\end{gather}
leaving a logarithmic divergence for both the planar and the non-planar part.
Due to symmetry reasons this result is also equal to the according correction to the $\bB\bB$ propagator, i.e.
\begin{gather*}
 \Sigma_{\m1\m2,\n1\n2}^{\rm \bB\bB}(p)=\Sigma_{\m1\m2,\n1\n2}^{\text{BB}}(p).
\end{gather*}

\subsection[Corrections to the $B\bB$ propagator]{Corrections to the $\boldsymbol{B\bB}$ propagator}
\label{sec:1loop_BbB}

\begin{figure}[t]
 \centering
 \includegraphics[scale=0.8]{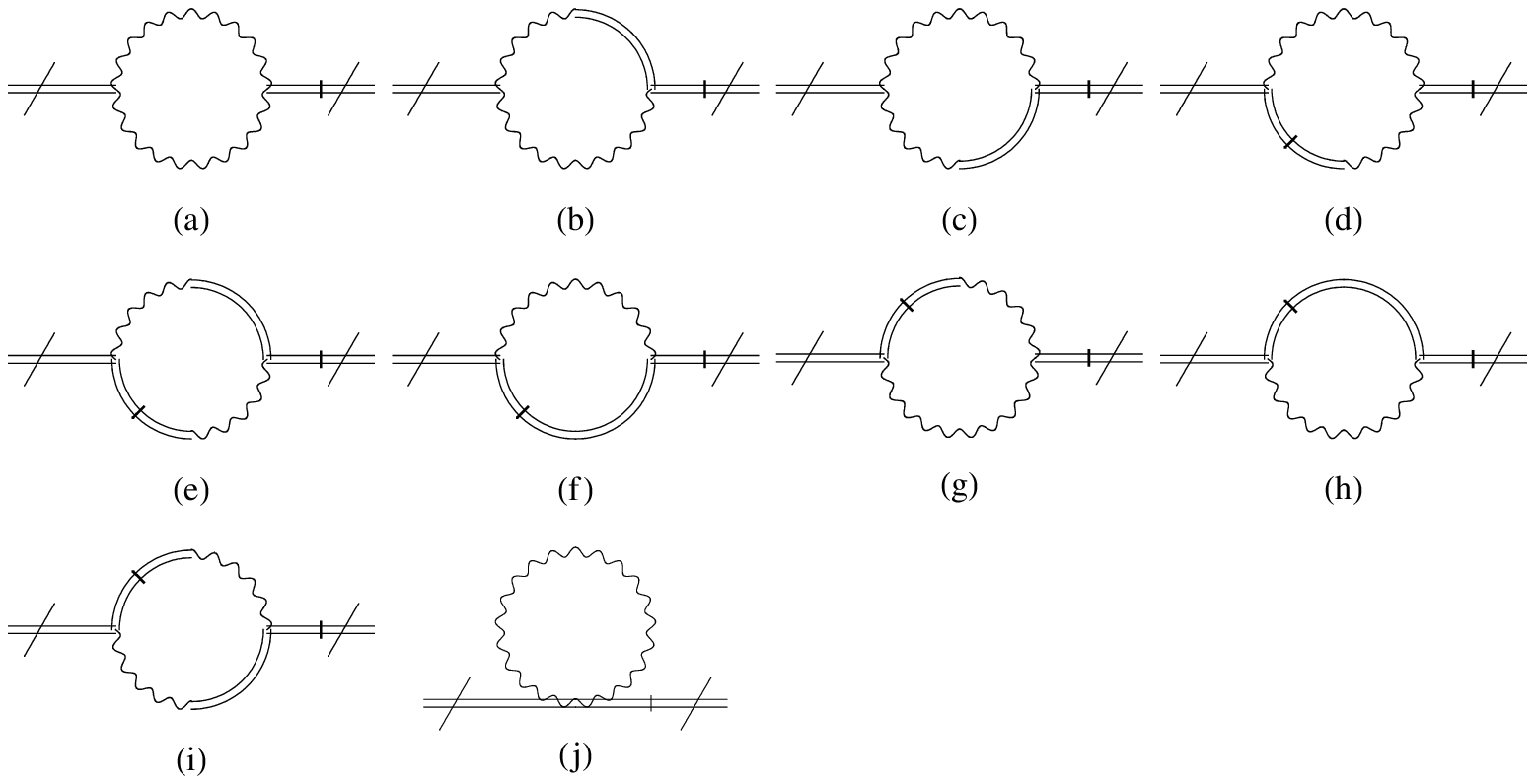}
 \caption{One loop corrections for $\langle   B_{\m_1\m_2} \bB_{\n_1\n_2}  \rangle$ (with amputated external legs).}
 \label{fig:1loop_BbB}
\end{figure}

\begin{table}[t]
 \centering
 \caption{Symmetry factors for the graphs depicted in \figref{fig:1loop_BbB}.}
 \vspace{1mm}

 \begin{tabular}{l c | l c | l c}
 \hline
 \hline
$(a)$ & 1/2 &  $(e)$ & 1 &  $(i)$ & 1\tsep{2pt}\\
$(b)$ & 1 &  $(f)$ & 1 &  $(j)$ & 1/2\\
$(c)$ & 1 &  $(g)$ & 1 &  & \\
$(d)$ & 1 &  $(h)$ & 1 &  & \\
\hline
\hline
 \end{tabular}
 \label{tab:1loop_BbB_div}
\end{table}

For the correction to $\langle B_{\m1\m2}\bB_{\n1\n2} \rangle$ one f\/inds the ten divergent graphs depicted in \figref{fig:1loop_BbB}.
Expansion for small external momenta $p$ and summation of the integrated results yields
\begin{gather*}
\Sigma_{\m1\m2,\n1 \n2}^{{\rm p,B\bB}}(p)= \frac{g^2}{2 \pi^2} \L ^2 \mu ^2 \p^2\left(\d_{\mu2\nu1} \d _{\mu 1 \n2}-\d _{\mu 1 \n1} \d _{\mu 2 \n2}\right)\nonumber\\
\phantom{\Sigma_{\m1\m2,\n1 \n2}^{{\rm p,B\bB}}(p)=}{}
+\frac{g^2 \l^2}{32 \pi^2} \left(\d _{\m1 \n1} \d_{\m2 \n2}-\d _{\m2 \n1} \d _{\m1 \n2}\right)  \mathop{{\rm K}_0}\left(2 \sqrt{\frac{M^2}{\L ^2}}\right)+{\rm f\/inite}, \nonumber\\
\Sigma_{\m1\m2,\n1 \n2}^{{\rm np,B\bB}}(p)= \frac{g^2 \l^2}{64 \pi^2} \bigg(\frac{\d_{\m1\n2} \p_{\m2}\p_{\n1}-\d_{\m1 \n1}\p_{\m2}\p_{\n2}-\d_{\mu 2 \n 2} \p_{\m1} \p_{\n1}+\d_{\m2\n1}\p_{\m1}\p_{\n2}}{\p^2}\nonumber\\
 \phantom{\Sigma_{\m1\m2,\n1 \n2}^{{\rm np,B\bB}}(p)=}{}
 +2  \mathop{{\rm K}_0}\left(\sqrt{M^2 \p^2}\right) \left(\d_{\m1 \n2} \d_{\m2 \n1}-\d_{\m1 \n1} \d _{\m2 \n2}\right)\bigg)+{\rm f\/inite} ,
\end{gather*}
Hence, the divergent part is given by
\begin{gather*}
\Sigma_{\m1\m2,\n1\n2}^{\text{B}\bB}(p)=\frac{g^2}{2 \pi^2} \L^2 \mu ^2 \p^2\left(\d_{\mu2\nu1} \d_{\mu 1 \n2}-\d_{\mu 1 \n1} \d_{\mu 2 \n2}\right)\nonumber\\
\phantom{\Sigma_{\m1\m2,\n1\n2}^{\text{B}\bB}(p)=}{} +\frac{g^2 \l^2}{64 \pi^2}\left(\d_{\m1 \n1} \d_{\m2 \n2}-\d_{\m2 \n1} \d_{\m1 \n2}\right)\left(\ln\L^2+\ln\p^2\right)+{\rm f\/inite},
\end{gather*}
which is logarithmically divergent in $\p^2$ and quadratically in $\Lambda$. Once more, $M$ has dropped out in the sum of planar and non-planar contributions.
Furthermore, note that $\Sigma_{\m1\m2,\n1\n2}^{\rm B\bB}\equiv\Sigma_{\n1\n2,\m1\m2}^{\rm \bB B}$ as is obvious from the result \eqref{eq:1loop_BB_final}.

\subsection{Dressed gauge boson propagator and analysis}
\label{sec:1loop_analysis}

In the standard renormalization procedure, the dressed propagator at one-loop level is given by
\begin{gather}\label{eq:dressed1}
\raisebox{-11pt}[0pt][0pt]{\includegraphics[scale=0.23]{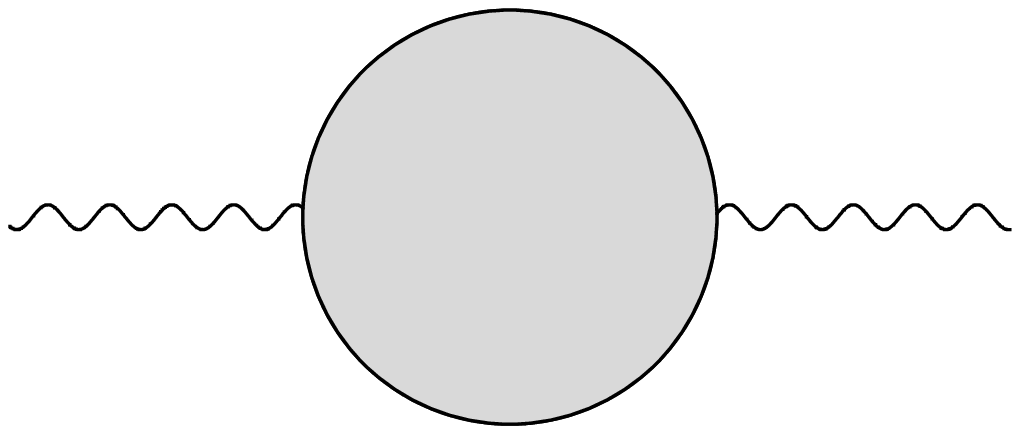}}
\equiv \Delta'(p) = \inv{\mathcal A} + \inv{\mathcal A}
\Sigma(\Lambda,p)  \inv{\mathcal A}  ,
\end{gather}
where
\begin{gather*}
  \inv{\mathcal A} \equiv  G^{\rm AA}_{\m\n}(p) ,
\qquad
 \Sigma(\Lambda,p)\equiv \big( \Pi^{\text{plan}}
\big)_{{\rm regul.}} (\Lambda,p) +\Pi^{\text{n-pl}}(p) .
\end{gather*}
For $\mathcal A\neq 0$, one can apply the formula
\begin{gather}
\label{eq:Expansion}
\inv{\mathcal A+\mathcal B}=\inv{\mathcal A}-\inv{\mathcal A}\, \mathcal B \, \inv{\mathcal A+\mathcal B} = \inv{\mathcal A}-\inv{\mathcal A}\, \mathcal B
\, \inv{\mathcal A} + {\mathcal O} (\mathcal B^2)   ,
\end{gather}
which allows one to rewrite expression \eqref{eq:dressed1} to order
$\Sigma$ as
\begin{gather*}
\Delta'(p) =\inv{\mathcal A-\Sigma(\Lambda,p)}  ,
\end{gather*}
and thus (in the case of renormalizability) to absorb any divergences in the appropriate parameters of the theory present in $\mathcal A$ (see \cite{Blaschke:2008b} for an example).

However, in our case \eqref{eq:Expansion} cannot be applied directly, as the complete one loop correction to the gauge boson propagator is given by the sum of all the results of Sections~\ref{sec:1loop_vacpol}--\ref{sec:1loop_BbB} after multiplication with {\em appropriate, i.e.\ different} external legs:
\begin{gather}
G^{\rm AA,1l-ren}_{\m\n}(p)=G^{\rm AA}_{\m\n}(p)+G^{\rm AA}_{\m\r}(p)\Pi_{\r\s}(p)G^{\rm AA}_{\s\n}(p)
  + G^{\rm AA}_{\m\r}(p)2\Sigma^{\rm AB}_{\r,\s1\s2}(p)G^{\rm BA}_{\s1\s2,\n}(p)\nonumber\\
\phantom{G^{\rm AA,1l-ren}_{\m\n}(p)=}{}  + G^{\rm AA}_{\m\r}(p)2\Sigma^{\rm A\bB}_{\r,\s1\s2}(p)G^{\rm \bB A}_{\s1\s2,\n}(p)
  + G^{\rm AB}_{\m,\r1\r2}(p)\Sigma^{\rm BB}_{\r1\r2,\s1\s2}(p)G^{\rm BA}_{\s1\s2,\n}(p)\nonumber\\
\phantom{G^{\rm AA,1l-ren}_{\m\n}(p)=}{}  + G^{\rm AB}_{\m,\r1\r2}(p)2\Sigma^{\rm B\bB}_{\r1\r2,\s1\s2}(p)G^{\rm \bB A}_{\s1\s2,\n}(p)
  + G^{\rm A\bB}_{\m,\r1\r2}(p)\Sigma^{\rm \bB\bB}_{\r1\r2,\s1\s2}(p)G^{\rm \bB A}_{\s1\s2,\n}(p)\nonumber\\
 \phantom{G^{\rm AA,1l-ren}_{\m\n}(p)=}{}
  +\mathcal{O}\left(g^4\right).\label{eq:DressedPropComplete1}
\end{gather}
Note, that the factors 2 stem from the (not explicitly written) mirrored contributions $AB\leftrightarrow BA$, $A\bB\leftrightarrow \bB A$, and $B\bB\leftrightarrow \bB B$.
Since the factor $\mathcal A$ must be the same for all summands we have to use the Ward Identities
\eqref{prop-rel_a} and \eqref{prop-rel_c}, i.e.
\begin{gather}
 G^{AB}_{\m,\r\s}(k) =G^{A\bB}_{\m,\r\s}(k)=-G^{BA}_{\r\s,\m}(k)=-G^{\bB A}_{\r\s,\m}(k),\nonumber\\
 2k^2\k^2 G^{AB}_{\r,\m\n}(k) =\ri\frac{a'}{\m}\left(k_\m G^{AA}_{\r\n}(k)-k_\n G^{AA}_{\r\m}(k)\right),\label{eq:ABtoAA}
\end{gather}
which allow us to express the (tree level) $AB$ and $A\bar{B}$ propagators uniquely in terms of $AA$-propagators.
This leads (in analogy to \eqref{eq:Expansion}) to the following representation for the dressed one-loop gauge boson propagator:
\begin{gather*}
G^{\rm AA,1l-ren}_{\m\n}(p)=\inv{\mathcal A}-\inv{\mathcal A} \Big(\sum \mathcal B_i\Big) \inv{\mathcal A},
\end{gather*}
where $1/\mathcal A$ once more stands for the tree level gauge boson propagator. The $\mathcal B_i$'s are given by the one-loop corrections (with amputated external legs) of the two-point functions relevant for the dressed gauge boson propagator, multiplied by any prefactors coming from \eqref{eq:ABtoAA} and the factor $2$ where needed (c.f.~\eqref{eq:DressedPropComplete1}). Thus, the full propagator is given by
\begin{gather*}
 G^{\rm AA,1l-ren}_{\m\n}(p)=G^{\rm AA}_{\m\n}(p)+G^{\rm AA}_{\m\r}(p)\Pi_{\r\s}(p)G^{\rm AA}_{\s\n}(p)\nonumber\\
 \quad{} + \left(\frac{\ri a'}{\m p^2\p^2} \right)\Bigg\{2G^{\rm AA}_{\m\r}(p)\left(\Sigma^{\rm AB}_{\r,\s1\s2}(p)+\Sigma^{\rm A\bar{B}}_{\r,\s1\s2}(p)\right)p_{\s2}G^{\rm AA}_{\n\s1}(p) \\ 
 \quad{}+ \left(\tfrac{\ri a'}{\m p^2\p^2} \right) p_{\r1}G^{\rm AA}_{\m\r2}(p)
\left(\Sigma^{\rm BB}_{\r1\r2,\s1\s2}(p)+2\Sigma^{\rm B\bB}_{\r1\r2,\s1\s2}(p)+\Sigma^{\rm \bB\bB}_{\r1\r2,\s1\s2}(p)\right)p_{\s2}G^{\rm AA}_{\n\s1}(p)\Bigg\}.\nonumber
\end{gather*}
The expression $\mathcal{B}=\sum\limits_i\mathcal{B}_i$ for $M\to0$ is explicitly given by
\begin{gather*}
 \mathcal{B}= \frac{g^2}{8\pi^2\mu^4} \Bigg\{\p_\m\p_\n\left(\frac{16\mu^4}{(\p^2)^2}+\frac{\theta^4\lambda^4}{2(\p^2)^4}\right)-7\l^2\m^2\frac{\theta^4}{(\p^2)^4}\left(p^2\d_{\m\n}-p_\m p_\n\right)\left(4-\p^2\Lambda^2\right)\nonumber\\
\phantom{\mathcal{B}=}{}
+ \left(p^2 \delta _{\m \n}-p_{\m} p_{\n}\right) \left[\ln2-\ln|\p|-\ln\L\right]\left(\frac{5}{3} \mu^4+\frac{3\lambda^2 \mu^2\theta^2}{(\p^2)^2}+\frac{\lambda^4\theta^4}{(\p^2)^4}\right)\Bigg\}
 +{\rm f\/inite},
\end{gather*}
and shows us two things: In contrast to commutative gauge models and even though the vacuum polarization tensor $\Pi_{\m\n}$ only had a logarithmic UV divergence, the full $\mathcal{B}$ diverges quadratically in the UV cutof\/f $\L$. Secondly, despite the fact that $\Pi_{\m\n}$ exhibited the usual quadratic IR divergence, $\mathcal B$ behaves like $\inv{(\p^2)^3}$ in the IR limit. Both properties arise due to the existence (and the form) of the mixed $AB$ and $A\bB$ propagators, and seem problematic concerning renormalization for two reasons: On the one hand, the form of the propagator is modif\/ied implying new counter terms in the ef\/fective action. On the other hand, higher loop insertions of this expression can lead to IR divergent integrals, as will be discussed in the next section.

\section{Higher loop calculations}
\label{sec:higher_loop_calc}

In the light of higher loop calculations it is important to investigate the IR behaviour of expected integrands with insertions of the one-loop corrections being discussed in \secref{sec:one_loop}. The aim is to identify possible poles at $\p^2=0$. Hence, we consider a chain of $n$ non-planar insertions denoted by $\Xi^{\phi_1\phi_2}(p,n)$, which may be part of a higher loop graph. Every insertion $\Xi$ represents the sum of all divergent one-loop contributions with external f\/ields $\phi_1$ and $\phi_2$ (cf. Sections~\ref{sec:1loop_vacpol}--\ref{sec:1loop_BbB}). Due to the numerous possibilities of constructing such graphs, we will examine only a few exemplary conf\/igurations in this section~-- especially those for which one expects the worst IR behaviour.

\begin{figure}[t]
  \centering
 \includegraphics[scale=0.64]{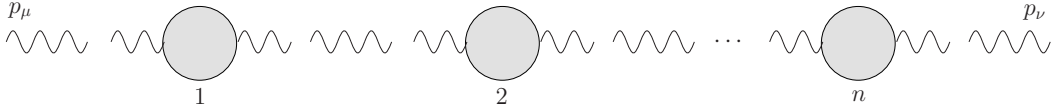}
 \caption{A chain of $n$ non-planar insertions, concatenated by gauge f\/ield propagators.}
 \label{fig:nloop_AA}
\end{figure}

To start with, let us state that amongst all types of two point functions, the vacuum polarization shows the highest, namely a quadratic divergence. Amongst the propagators those with two external double-indexed legs, e.g.~$B$ or~$\bB$ feature the highest (quartic) divergence in the limit of vanishing external momenta. A chain of $n$ vacuum polarizations $\Pi_{\m\n}^{\text{np}}(p)$ (see equations~\eqref{eq:1loop_AA_o0_np} and \eqref{eq:1loop_AA_o2_np}) with $(n+1)$ $AA$-propagators ($(n-1)$ between the individual vacuum polarization graphs, and one at each end) leads to the following expression (for a graphical representation, see \figref{fig:nloop_AA}):
\begin{gather*}
\Xi^{A A}_{\m\n}(p,n) =\left(G^{AA}(p)\Pi^{\text{np}}(p)\right)^n_{\m\r}G^{AA}_{\r\n}(p)
 =\left(\frac{2g^2}{\pi^2}\right)^n\frac{1}{\left(p^2+\frac{a'^2}{\p^2}\right)^{n+1}}
 \frac{\p_{\m}\p_{\n}}{(\p^2)^{n+1}} .
\end{gather*}
Note that due to transversality, from the propagator \eqref{eq:prop_aa} only the term with the Kronecker delta enters the calculation.
For vanishing momenta, i.e.\ in the limit $\p^2\to0$ the expression reduces to
\begin{gather*}
\lim\limits_{\p^2\to0}\Xi^{A A}_{\m\n}(p,n)=\left(\frac{2g^2}{\pi^2}\right)^n\frac{\p_{\m}\p_{\n}}{a'^{2(n+1)}}  ,
\end{gather*}
exhibiting IR f\/initeness which is independent from the number of inserted loops.

Another representative is the chain
 \[\Xi^{A \phi}(p,n)\equiv G^{A\phi}(p)\left(\Sigma^{\rm np,\phi A}(p)G^{A\phi}(p)\right)^n, \qquad\text{where}\quad\phi\in\{B,\bB\},
 \]
 which could replace any single $G^{A B}$ (or $G^{A \bB}$) line. Obviously, one has
\begin{gather*}
 \Xi^{A\phi}_{\m,\n1\n2}(p,n)= \frac{\ri a'}{2\m}\left(-\frac{3 g^2}{32 \pi^2} a'^2\right)^n \frac{\left(p_{\n1}\d_{\m\,\n2}-p_{\n2}\d_{\m\,\n1}\right)}{p^2\left[\p^2\left(p^2+\frac{a'^2}{\p^2}\right)\right]^{n+1}} n \ln\p^2 ,
 \end{gather*}
 which for $\p^2\ll1$ (and neglecting dimensionless prefactors) behaves like
 \begin{gather*}
\Xi^{A\phi}_{\m,\n1\n2}(p,n)\approx n\frac{\left(p_{\n1}\d_{\m\,\n2}-p_{\n2}\d_{\m\,\n1}\right)}{\m p^2}\ln \p^2.
\end{gather*}
The latter insertion can be regularized since the pole at $p=0$ is independent of $n$. In contrast, higher divergences are expected for chain graphs being concatenated by propagators with four indices, i.e.\ $G^{\bB B}_{\m\n,\r\s}$, $G^{B B}_{\m\n,\r\s}$, $G^{\bpsi \psi}_{\m\n,\r\s}$, due to the inherent quartic IR singularities. Let us start with the combination $\Xi^{\bB B}(p,n)\equiv\big(G^{\bB B}(p)\Sigma^{{\rm p,B\bB}}(p)\big)^nG^{\bB B}(p)$. As before, we can approximate for $\p^2\ll1$ and, omitting dimensionless prefactors and indices, f\/ind
\begin{gather*}
 \Xi^{A \phi}(p,n)\underset{\p^2\ll 1}{\propto}\frac{n}{\m^2}\frac{\ln\p^2}{\left(p^2\p^2\right)^n},
\end{gather*}
which represents a singularity $\forall\,n>1$ (since in any graph, at $n=0$, the divergence is regularized by the phase factor being a sine function which behaves like $p$ for small momenta). Regarding the index structures, no cancellations can be expected since the product of an arbitrary number of contracted, completely antisymmetric tensors is again an antisymmetric tensor with the outermost indices of the chain being free.

Exactly the same result is obtained for $\Xi^{B B}(p)\equiv\left(G^{B B}(p)\Sigma^{{\rm p,B B}}(p)\right)^nG^{B B}(p)$. From this it is clear that the damping mechanism seen in $\Xi^{AA}(p,n)$ fails for higher insertions of $B/\bB$ (and also $\psi/\bpsi)$ f\/ields).

\section{Discussion}
\label{sec:discussion}

We have elaborated on our recently introduced {\nc} gauge model~\cite{Blaschke:2009b}. Initially, the intent was to apply Algebraic Renormalization (AR), as was suggested by Vilar {\etal}~\cite{Vilar:2009}. In the light of that renormalization scheme it is most important to maximise the symmetry content of the theory which is the basis for the generation of constraints to potential counter terms. Therefore, after recapitulating general properties of our model, we studied the resulting algebra of symmetries.
However, as we exposed recently~\cite{Blaschke:2009c}, the foundations of AR are only proved to be valid in \emph{local} QFTs so far, and hence may not be applicable in {\nc} f\/ield theories, as the deformation inherently implies non-locality. In order to f\/ind a way out of this dilemma, explicit loop-calculations were presented, and our hope was to show renormalizability~-- at least at the one-loop level. In this respect, unexpected dif\/f\/iculties appeared. The soft breaking term, being required to implement the IR damping behaviour of the $1/p^2$ model in a way being compatible with the Quantum Action Principle of AR, gives rise to mixed propagators~$G^{AB}$ and~$G^{A\bB}$. These, in turn, allow the insertion of one-loop corrections with external $B$-f\/ields into the dressed $AA$ propagator (see \secref{sec:higher_loop_calc}) and, therefore, enter the renormalization. Despite all corrections featuring the expected~$\inv{\p^2}$ IR behaviour, the dressed propagators with external~$AB$ or~$A\bB$ legs multiplicatively receive higher poles due to the inherent quadratic divergences in $G^{AB}(p)$ (and~$G^{A\bB}(p)$) for $p\to 0$.
As a consequence, the resulting corrections cannot be absorbed in a straightforward manner.

However, renormalizability of the non-local model \eqref{eq:act_old_complete} cannot depend on how it is localized due to equivalence of the respective path integrals (see~\cite{Blaschke:2009b}). Therefore, we expect the same problems to appear in \emph{all} localized versions of \eqref{eq:act_old_complete}, including the one of Vilar {\etal}~\cite{Vilar:2009}. In fact, from the discussion in Appendix~\ref{app:vilar}, one notices that the propagators \eqref{eq:vilar_prop-bBbB}--\eqref{eq:vilar_prop-BbB} and \eqref{eq:vilar_prop-xibxi} of their action all exhibit the same quartic IR divergences as those of our present model~\eqref{eq:act_complete}, even though the operator $D_\m$ appears at most quadratically as $D^2$ in the according action \eqref{eq:vilar-action}. Nonetheless, the authors claim to have shown renormalizability using Algebraic Renormalization, which as we have discussed in~\cite{Blaschke:2009c} may not be applicable in {\nc} theories.

In this respect it has to be noted that in commutative space the model of Vilar {\etal}~\cite{Vilar:2009} should indeed be renormalizable, since the action, apart from the star product, is completely local and provides the necessary symmetries for the Quantum Action Principle. Since the propagators are the same in both spaces, and hence show the same quartic IR divergences, one may expect related IR problems to cancel when considering the sum of bosonic and fermionic sectors (i.e.~$B/\chi$ and~$\psi/\xi$). These cancellations should also take place in {\nc} space (in both models), but the problem of proving renormalization remains (cf.\ \secref{sec:1loop_analysis}).

Coming back to the problem of IR divergent propagators we have also investigated the structure of singularities in higher-loop integrands by studying chain graphs consisting of interleaving tree-level propagators, and one-loop corrections of various types. It turned out that chains containing gauge f\/ields benef\/it from the damping of the propagator \eqref{eq:prop_aa} while those consisting (solely) of concatenated $B$ and $\bB$ f\/ields and insertions do (expectedly) not. Hence, at f\/irst sight, there exist divergences which increase order by order, which would indicate non-renormalizability. However, we may point out that, due to the symmetry between the $B/\bB$ and $\psi/\bpsi$ sectors, cancellations can be expected. These already appear in our one-loop calculations, and there is strong evidence that they appear to all orders. An intuitive argument can be given when considering the action~\eqref{act-loc} for $\l\to0$, i.e.\ vanishing damping. In this case, the $B/\bB$ and $\psi/\bpsi$ f\/ields may simply be integrated out in the path integral formalism (see~\cite{Blaschke:2009b}), and the contributions cancel exactly.
An alternative approach which avoids these uncertainties is in preparation.

\appendix

\section{Vertices}
\label{app:vertices}

\vspace*{-5mm}

\begin{gather*}
 \raisebox{-20pt}[0pt][0pt]{\includegraphics[scale=0.8]{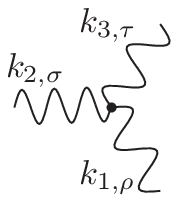}}  =\widetilde{V}^{3A}_{\rho\s\tau}(k_1, k_2, k_3)
 =2\ig(2\pi)^4\d^4(k_1+k_2+k_3)\sin\left(\tfrac{k_1\k_2}{2}\right) \nonumber\\
\hspace*{19mm}{}\times\left[(k_3-k_2)_\rho \d_{\s\tau}+(k_1-k_3)_\s \d_{\rho\tau}+(k_2-k_1)_\tau \d_{\rho\s}\right],\\[2ex]
\raisebox{-18pt}[20pt][0pt]{\includegraphics[scale=0.8]{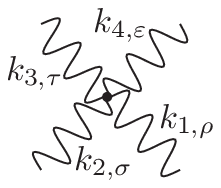}}
=\widetilde{V}^{4A}_{\rho\s\tau\e}(k_1, k_2, k_3, k_4) =-4g^2(2\pi)^4\d^4(k_1+k_2+k_3+k_4) \nonumber\\
\hspace*{22.5mm}{}\times\left[(\d_{\rho\tau}\d_{\s\e}-\d_{\rho\e}\d_{\s\tau})\sin\left(\tfrac{k_1\k_2}{2}\right)
\sin\left(\tfrac{k_3\k_4}{2}\right)\right. \nonumber\\
\hspace*{22.5mm}{}
 +(\d_{\rho\s}\d_{\tau\e}-\d_{\rho\e}\d_{\s\tau})\sin\left(\tfrac{k_1\k_3}{2}\right)
 \sin\left(\tfrac{k_2\k_4}{2}\right)\nonumber\\
 \left.
\hspace*{22.5mm}{} +(\d_{\rho\s}\d_{\tau\e}-\d_{\rho\tau}\d_{\s\e})\sin
\left(\tfrac{k_2\k_3}{2}\right)\sin\left(\tfrac{k_1\k_4}{2}\right)\right],
\\[1ex]
 \raisebox{-19pt}[0pt][0pt]{\includegraphics[scale=0.8]{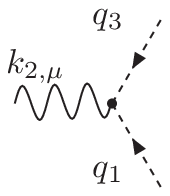}} =\widetilde{V}^{\bc Ac}_\mu(q_1, k_2, q_3)
 =-2\ig(2\pi)^4\d^4(q_1+q_2+k_3)q_{1\mu}\sin\left(\tfrac{q_1\q_3}{2}\right),
\\[3.5ex]
\raisebox{-18pt}[20pt][0pt]{\includegraphics[scale=0.8]{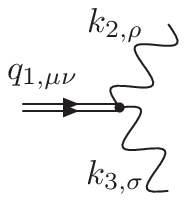}}=\widetilde{V}^{BAA}_{\mu\nu,\rho\s}(q_1, k_2, k_3)=\hspace{1mm} \raisebox{-18pt}[0pt][0pt]{\includegraphics[scale=0.8]{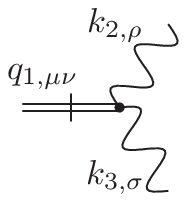}}\widetilde{V}^{\bB AA}_{\mu\nu,\rho\s}(q_1, k_2, k_3)\nonumber\\[2.5ex]
\hspace*{14.5mm}{}=\l g (2\pi)^4\d^4(q_1+k_2+k_3)\left(\d_{\mu\rho}\d_{\nu\s}-\d_{\mu\s}\d_{\nu\rho}\right)\sin\left(\tfrac{k_2\k_3}{2}\right),
\\
\raisebox{-18pt}{\includegraphics[scale=0.8]{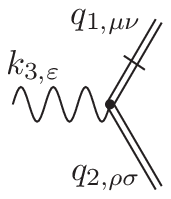}} =\widetilde{V}^{\bB BA}_{\mu\nu,\rho\s\e}(q_1, q_2, k_3)=
-\raisebox{-18pt}{\includegraphics[scale=0.8]{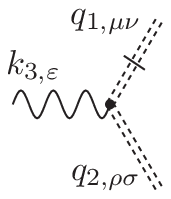}}=-\widetilde{V}^{\bpsi \psi A}_{\mu\nu,\rho\s\e}(q_1, q_2, k_3)\nonumber\\
\hspace*{13.5mm}{} =-\ri \mu^2 g(2\pi)^4\d^4(q_1+q_2+k_3)\left(\d_{\mu\rho}\d_{\nu\s}-\d_{\mu\s}\d_{\nu\rho}\right) \nonumber\\
\hspace*{18mm}{} \times\left((\q_1)^2+(\q_2)^2\right)\left(q_1-q_2\right)_\e\sin\left(\tfrac{q_1\q_2}{2}\right),
\\
\raisebox{-18pt}{\includegraphics[scale=0.8]{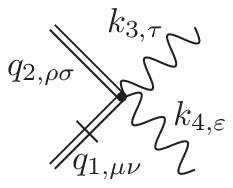}} =\widetilde{V}^{\bB B2A}_{\mu\nu,\rho\s,\tau\e}(q_1, q_2, k_3,k_4)=\raisebox{-18pt}{\includegraphics[scale=0.8]{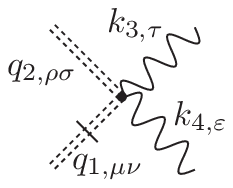}}=-\widetilde{V}^{\bpsi \psi 2A}_{\mu\nu,\rho\s,\tau\e}(q_1, q_2, k_3,k_4)\nonumber\\
\hspace*{18.5mm}{} =2\mu^2g^2\th^2(2\pi)^4\d^4(q_1+q_2+k_3+k_4)\left(\d_{\mu\rho}\d_{\nu\s}-\d_{\mu\s}\d_{\nu\rho}\right) \\
\hspace*{18.5mm}{}\times \bigg\{\left[k_{3,\tau}k_{4,\e}\!+2\!\left(q_{1,\tau}k_{4,\e}\!+q_{2,\e}k_{3,\tau}\right)\!
+4q_{1,\tau}q_{2,\e}\!-\d_{\e\tau}\!\left({q_1}^2\!+{q_2}^2\right)\right]\!
\sin\!\big(\tfrac{q_1\k_3}{2}\!\big)\!\sin\!\big(\tfrac{q_2\k_4}{2}\!\big)\nonumber\\
\hspace*{18.5mm}{}+\left[k_{3,\tau}k_{4,\e}\!+2\!\left(q_{2,\tau}k_{4,\e}\!+q_{1,\e}k_{3,\tau}\right)\!
+4q_{1,\e}q_{2,\tau}\!-\d_{\e\tau}\!\left({q_1}^2\!+{q_2}^2\right)\right]
\!\sin\!\big(\tfrac{q_1\k_4}{2}\!\big)\!\sin\!\big(\tfrac{q_2\k_3}{2}\!\big)\!\bigg\},\nonumber
\\
\raisebox{-18pt}{\includegraphics[scale=0.8]{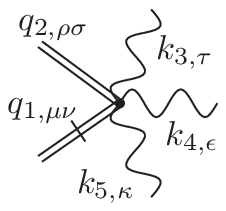}} =\widetilde{V}^{\bB B3A}_{\mu\nu,\rho\s,\tau\e\kappa}\,(q_1, q_2, k_3,k_4,k_5)=-\raisebox{-18pt}{\includegraphics[scale=0.8]{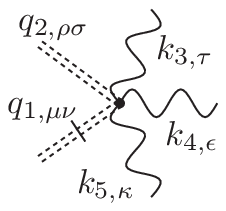}} =-\widetilde{V}^{\bpsi \psi 3A}_{\mu\nu,\rho\s,\tau\e\kappa}\,(q_1, q_2, k_3,k_4,k_5)\nonumber\\
 \hspace*{17.5mm}{}=-4\ig^3\mu^2\th^2(2\pi)^4\d^4(q_1+q_2+k_3+k_4+k_5)\left(\d_{\mu\rho}\d_{\nu\s}-\d_{\mu\s}\d_{\nu\rho}\right)\times\nonumber\\
\hspace*{17.5mm}{} \times\Bigg\{\left[k_{3}+2q_{1}\right]_\tau\d_{\e\kappa}\sin\!\big(\tfrac{k_3\q_1}{2}\!\big)\Bigg[\sin\!\big(\tfrac{k_5\q_2}{2}\!\big)\sin\!\big(\tfrac{k_4(\k_5+\q_2)}{2}\!\big)+(k_4\leftrightarrow k_5)\Bigg]\nonumber\\
 \hspace*{17.5mm}{}+\left[k_4+2q_{1}\right]_\e\d_{\tau\kappa}\sin\!\big(\tfrac{k_4\q_1}{2}\!\big)\Bigg[\sin\!\big(\tfrac{k_5\q_2}{2}\!\big)\sin\!\big(\tfrac{k_3(\k_5+\q_2)}{2}\!\big)+(k_5\leftrightarrow k_3)\Bigg]\\ 
\hspace*{17.5mm}{} +\left[k_5+2q_{1}\right]_\kappa\d_{\tau\e}\sin\!\big(\tfrac{k_5\q_1}{2}\!\big)\Bigg[\sin\!\big(\tfrac{k_3\q_2}{2}\!\big)
 \sin\!\big(\tfrac{k_4(\k_3+\q_2)}{2}\!\big)+(k_3\leftrightarrow k_4)\Bigg]\! +(q_1\leftrightarrow q_2)\Bigg\},
\nonumber \\
\raisebox{-18pt}{\includegraphics[scale=0.8]{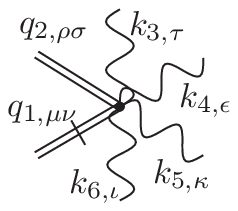}} =\widetilde{V}^{\bB B4A}_{\mu\nu,\rho\s,\tau\e\kappa\iota}\,(q_1, q_2, k_3,k_4,k_5,k_6)=-\raisebox{-18pt}{\includegraphics[scale=0.8]{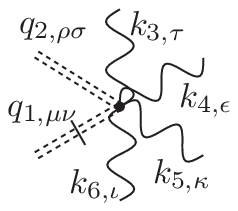}} =-\widetilde{V}^{\bpsi \psi 4A}_{\mu\nu,\rho\s,\tau\e\kappa\iota}\,(q_1, q_2, k_{3-6})\nonumber\\
 \hspace*{18.5mm}{} =2g^4\mu^2\th^2(2\pi)^4\d^4(q_1+q_2+k_3+k_4+k_5+k_6)\left(\d_{\mu\rho}\d_{\nu\s}-\d_{\mu\s}\d_{\nu\rho}\right) \nonumber\\
\hspace*{18.5mm}{} \times\!\Bigg\{\!2\d_{\tau\e}\d_{\kappa\iota}\!\Big[\!\sin\!\big(\tfrac{k_4\q_1}{2}\big)\!\sin\!\big(\tfrac{k_3(\k_4\!+\!\q_1)}{2}\big)
\sin\!\big(\tfrac{k_6\q_2}{2}\big)\sin\!\big(\tfrac{k_5(\k_6\!+\!\q_2)}{2}\big)\!+\!(k_3\!\leftrightarrow \!k_4)\!+\!(k_5\!\leftrightarrow\! k_6)\!\Big]\nonumber\\
\hspace*{18.5mm}{} +\d_{\tau\kappa}\d_{\e\iota}\Big[\!\sin\!\big(\tfrac{k_5\q_1}{2}\!\big)\sin\!\big(\tfrac{k_3(\k_5+\q_1)}{2}\!\big)
 \sin\!\big(\tfrac{k_6\q_2}{2}\!\big)\sin\!\big(\tfrac{k_4(\k_6+\q_2)}{2}\!\big)\!+\!(k_3\leftrightarrow k_5)\!+\!(k_4\leftrightarrow k_6)\!\Big]\nonumber\\
 \hspace*{18.5mm}{}+\d_{\tau\iota}\d_{\kappa\e}\Big[\!\sin\!\big(\tfrac{k_6\q_1}{2}\!\big)\sin\!\big(\tfrac{k_3(\k_6+\q_1)}{2}\!\big)
 \sin\!\big(\tfrac{k_4\q_2}{2}\!\big)\sin\!\big(\tfrac{k_5(\k_4+\q_2)}{2}\!\big)\!+\!(k_3\leftrightarrow k_6)\!+\!(k_5\leftrightarrow k_4)\!\Big]\nonumber\\
\hspace*{18.5mm}{} +(q_1\leftrightarrow q_2)\Bigg\}.
\end{gather*}

\section{Propagators of the model by Vilar et al.}
\label{app:vilar}

The tree level action of~\cite{Vilar:2009} is given by:
\begin{gather}
 \Act  = \Act_{0} + \Act_{\text{break}} + \Act_{\text{G}} + \Act_{\text{gf}}, \nonumber  \\
\Act_{0}  = \intx \left[\inv{4}F_{\mu\nu}\star{F}^{\mu \nu }+ \bar{\chi}_{\mu\nu}\star D^{2} B^{\mu\nu}+\bB_{\mu\nu}\star D^{2} \chi^{\mu\nu}+\gamma^{2}\bar{\chi}_{\mu\nu}\star\chi^{\mu\nu} \right] ,\nonumber\\
\Act_{\text{break}}  = \intx \left[ -\ri\frac{\gamma}{2}B_{\mu\nu}\star{F}^{\mu\nu} +\ri\frac{\gamma}{2}\bB_{\mu\nu}\star{F}^{\mu\nu} \right] ,\nonumber\\
\Act_{\text{G}} =\intx\left[- \bpsi_{\mu\nu}\star D^2\star\xi^{\mu\nu} - \bar{\xi}_{\mu\nu}\star D^2\psi^{\mu\nu} -\gamma^{2}\bpsi_{\mu\nu}\star\psi^{\mu\nu}\right] ,\nonumber\\
\Act_{\text{gf}} =\intx\left[\ri b\star\partial^\mu A_\mu+\bc\star\partial^\mu D_\mu c\right] ,\label{eq:vilar-action}
\end{gather}
where the complex conjugated pairs ($B_{\m\n}$, $\bB_{\m\n}$), ($\chi_{\m\n}$, $\bar\chi_{\m\n}$) are bosonic auxiliary f\/ields of mass dimension 1, and ($\psi_{\m\n}$, $\bpsi_{\m\n}$), ($\xi_{\m\n}$, $\bar\xi_{\m\n}$) are their associated ghost f\/ields.
From the bilinear parts of this action one derives the following 19 propagators:
\begin{subequations}
\begin{gather}
G^A_{\m\n}(k) =\frac{-1}{\nov{k}}\left(\d_{\m\n}-\frac{k_\m k_\n}{k^2}\right) ,\\
G^{BA}_{\r,\s\t}(k) =\frac{-\gamma^3}{\nov{k}}\frac{\left(k_\s\d_{\r\t}-k_\t\d_{\r\s}\right)}{2(k^2)^2} ,\\
G^{\bB A}_{\r,\s\t}(k) =-G^{BA}_{\r,\s\t}(k) ,\\
G^{\chi A}_{\r,\s\t}(k) =\frac{\ri\gamma}{\nov{k}}\frac{\left(k_\s\d_{\r\t}-k_\t\d_{\r\s}\right)}{2k^2} ,\\
G^{\bar\chi A}_{\r,\s\t}(k) =-G^{\chi A}_{\r,\s\t}(k) ,\\
G^{\bB\bB}_{\r\s,\t\e}(k) =\frac{\gamma^4}{(k^2)^2}\frac{\left(k_\r k_\t\d_{\s\e}+k_\s k_\e\d_{\r\t}-k_\r k_\e\d_{\s\t}-k_\s k_\t\d_{\r\e}\right)}{4(k^2)^2\nov{k}} ,\label{eq:vilar_prop-bBbB}\\
G^{BB}_{\r\s,\t\e}(k) =G^{\bB\bB}_{\r\s,\t\e}(k) ,\label{eq:vilar_prop-BB}\\
G^{B\bB}_{\r\s,\t\e}(k) =\gamma^2\frac{\left(\d_{\r\t}\d_{\s\e}-\d_{\r\e}\d_{\s\t}\right)}{2(k^2)^2}
-G^{\bB\bB}_{\r\s,\t\e}(k) ,\label{eq:vilar_prop-BbB}\\
G^{\bar\chi\bar\chi}_{\r\s,\t\e}(k) =-\gamma^2\frac{\left(k_\r k_\t\d_{\s\e}+k_\s k_\e\d_{\r\t}-k_\r k_\e\d_{\s\t}-k_\s k_\t\d_{\r\e}\right)}{4(k^2)^2\nov{k}} ,\\
G^{\chi\chi}_{\r\s,\t\e}(k) =G^{\bar\chi\bar\chi}_{\r\s,\t\e}(k) ,\\
G^{\chi\bar\chi}_{\r\s,\t\e}(k) =-G^{\bar\chi\bar\chi}_{\r\s,\t\e}(k) ,\\
G^{\chi B}_{\r\s,\t\e}(k) =\frac{\gamma^4}{k^2}\frac{\left(k_\r k_\t\d_{\s\e}+k_\s k_\e\d_{\r\t}-k_\r k_\e\d_{\s\t}-k_\s k_\t\d_{\r\e}\right)}{4(k^2)^2\nov{k}} ,\\
G^{\bar\chi\bB}_{\r\s,\t\e}(k) =G^{\chi B}_{\r\s,\t\e}(k) ,\\
G^{\chi\bB}_{\r\s,\t\e}(k) =\frac{\left(\d_{\r\t}\d_{\s\e}-\d_{\r\e}\d_{\s\t}\right)}{2k^2}-G^{\chi B}_{\r\s,\t\e}(k) ,\\
G^{\bar\chi B}_{\r\s,\t\e}(k) =G^{\chi\bB}_{\r\s,\t\e}(k) ,
\\
G^{\bc c}(k) =-\inv{k^2} ,\\
G^{\xi,\bpsi}_{\m\n,\r\s}(k) =\frac{\left(\d_{\m\r}\d_{\n\s}-\d_{\m\s}\d_{\n\r}\right)}{2k^2} ,\\
G^{\bar\xi,\psi}_{\m\n,\r\s}(k) =-G^{\xi,\bpsi}_{\m\n,\r\s} ,\\
G^{\bar\xi\xi}_{\m\n,\r\s}(k) =-\gamma^2\frac{\left(\d_{\m\r}\d_{\n\s}-\d_{\m\s}\d_{\n\r}\right)}{2(k^2)^2} .\label{eq:vilar_prop-xibxi}
\end{gather}
\end{subequations}

\subsection*{Acknowledgements}

The authors are indebted to M.~Schweda and M.~Wohlgenannt for valuable discussions.
The work of D.N.~Blaschke, A.~Rofner and R.I.P.~Sedmik was supported by the ``Fonds zur F\"orderung der Wissenschaftlichen Forschung'' (FWF) under contract P20507-N16.

\pdfbookmark[1]{References}{ref}
\LastPageEnding

\end{document}